\documentclass[12pt,letterpaper]{article}
\usepackage[a4paper, total={7in, 10in}]{geometry}

\usepackage{graphicx}
\usepackage{helvet}
\usepackage{authblk}
\usepackage{hyperref}
\usepackage{amsmath} 
\usepackage{amssymb} 
\usepackage{orcidlink} 
\usepackage[super,comma,sort&compress]  
   {natbib}

\usepackage{graphicx}%
\usepackage{multirow}%
\usepackage{amsmath,amssymb,amsfonts}%
\usepackage{amsthm}%
\usepackage{mathrsfs}%
\usepackage[title]{appendix}%
\usepackage{xcolor}%
\usepackage{textcomp}%
\usepackage{manyfoot}%
\usepackage{booktabs}%
\usepackage{algorithm}%
\usepackage{algorithmicx}%
\usepackage{algpseudocode}%
\usepackage{listings}%

\usepackage[table]{xcolor} 
\definecolor{coolblue}{HTML}{D0E3FA} 
\definecolor{warmred}{HTML}{FAD4D4}  

\usepackage{xcolor} 

\definecolor{coolgreen}{RGB}{0, 150, 100}  
\newcommand{\adrian}[1]{\textcolor{black}{#1}}

\newcommand{\rev}[1]{\textcolor{black}{#1}}

\newcommand{\amr}[1]{\textcolor{black}{#1}}
\usepackage{multirow} 

\colorlet{red}{black}

\makeatletter
\renewcommand{\maketitle}{\bgroup\setlength{\parindent}{0pt}
\begin{flushleft}
  \textbf{\@title}
  
  \@author
\end{flushleft}\egroup}
\makeatother

\title{\fontsize{16}{18}\selectfont\bfseries Bacterial Gene Regulatory Neural Network as a Biocomputing Library of Mathematical Solvers}
\date{}

\author[1\orcidlink{0000-0003-1102-8095}, 4,  *]{Adrian  Ratwatte}
\author[1, 3]{Samitha Somathilaka}
\author[2, 3]{Thanh Cao}
\author[2, 3]{Xu Li}
\author[1, 3]{Sasitharan Balasubramaniam}

\affil[1]{School of Computing, University of Nebraska-Lincoln, 104 Schorr Center, 1100 T Street, Lincoln, 68588-0150, NE, USA}
\affil[2]{Civil and Environmental Engineering Department, University of
Nebraska-Lincoln, 844 N 16th St., C190, Scott Engineering Center, Lincoln, 68588-0531, NE, USA}

\affil[3]{These authors contributed equally}
\affil[4]{Lead contact}

\affil[*]{Correspondence: aratwatte2@huskers.unl.edu}

\begin{document}

\maketitle

\section*{SUMMARY}

Current biocomputing approaches predominantly rely on engineered circuits with fixed logic, offering limited stability and reliability under diverse environmental conditions. 
Here, we use the GRNN framework introduced in our previous work to transform bacterial gene expression dynamics into a biocomputing library of mathematical solvers. We introduce a sub-GRNN search algorithm that identifies functional subnetworks tailored to specific mathematical calculation  and classification tasks by evaluating gene expression patterns across chemically encoded input conditions. Tasks include identifying Fibonacci numbers, prime numbers, multiplication, and Collatz step counts. 
The identified problem-specific sub-GRNNs are then assessed using gene-wise and collective perturbation, as well as Lyapunov-based stability analysis, to evaluate robustness and reliability. 
Our results demonstrate that native transcriptional machinery can be harnessed to perform diverse mathematical calculation and classification tasks, while maintaining computing stability and reliability.

\section*{KEYWORDS}


Gene regulatory neural network (GRNN), Subnetwork search, Biocomputing, Perturbation analysis, Classification tasks, Mathematical calculation tasks, Lyapunov stability.

\section*{INTRODUCTION}


\amr{Recent advances in synthetic biology \cite{Goni, VELAZQUEZ2018415}, systems biology \cite{sharon2023trumpet}, and molecular engineering \cite{10654347, HUANG2022103163}} have reignited interest in biocomputing \cite{Li}  
alternative to silicon-based systems \cite{WADAN20258, Rizik2022, Bonnerjee2024MulticellularAN}. \amr{Living cells can be considered a unique computational substrate, characterized by massive parallelism \cite{doi:10.1073/pnas.79.8.2554, HasanBabu2025} and energy-efficient natural computing capabilities 
\cite{somathilaka2024wet, siuti2013synthetic, PIETAK2025112536}.} \adrian{These attributes are particularly attractive as traditional silicon technologies confront high energy demands, resource sustainability \cite{Gauthier} as well as challenges in extracting components to produce  processors 
~\cite{waldrop2016chips,markov2014limits, OTEROMURAS2023106836}.} Biocomputing frameworks that use endogenous \amr{cellular processes \cite{10464756, 7742378}}, such as transcriptional regulation and \amr{signal transduction \cite{11068111, 9122607}}, present new opportunities for developing self-contained, resource-aware, and robust computing platforms capable of operating within biological environments~\cite{grozinger2019pathways,elowitz2000synthetic, LI2023106315}.

Despite substantial advances in synthetic biology, current biocomputing platforms remain constrained by several key limitations. Most systems are built around engineered bacterial strains or designer cells that execute task-specific logic through hardwired gene circuits~\cite{Khalil2010,Nielsen2016,bonnerjee2024multicellular}. 
These implementations typically solve isolated arithmetic or logic problems using predefined input–output mappings and require manual rewiring \cite{Babazadeh} 
or extensive tuning to function reliably. As a result, they lack scalability and exploitation of 
native gene regulatory environment \cite{simpson2001whole}. 
\adrian{Furthermore, existing designs focus only on the final output and intermediate gene activity in computing, overlooking, the stability of the system}
\cite{jia2025dna}. 
\amr{Bridging these gaps requires biologically grounded computing frameworks that can generalize across tasks and use endogenous regulatory pathways \cite{8813106, balasubramaniam2023realizing} for stable and reliable computing.} 

Our previous work established the gene regulatory neural network (GRNN) framework by drawing parallels between gene regulation and perceptron-like computation in artificial neural networks~\cite{SOMATHILAKA2023100118, RATWATTE2024100158, balasubramaniam2023realizing}. 
Wet lab experiments were conducted with \textit{Escherichia coli strain K12}, where different chemical compounds representing unique numbers were added to the culture followed by retrieval of samples to obtain transcriptomic data. The aim is to determine the transcription factors that modulate gene expression through weighted regulatory interactions, which is analogous to weighted summation and activation in artificial neural networks. 

\begin{figure*}[t!] 
\centering
\includegraphics[width=\linewidth]{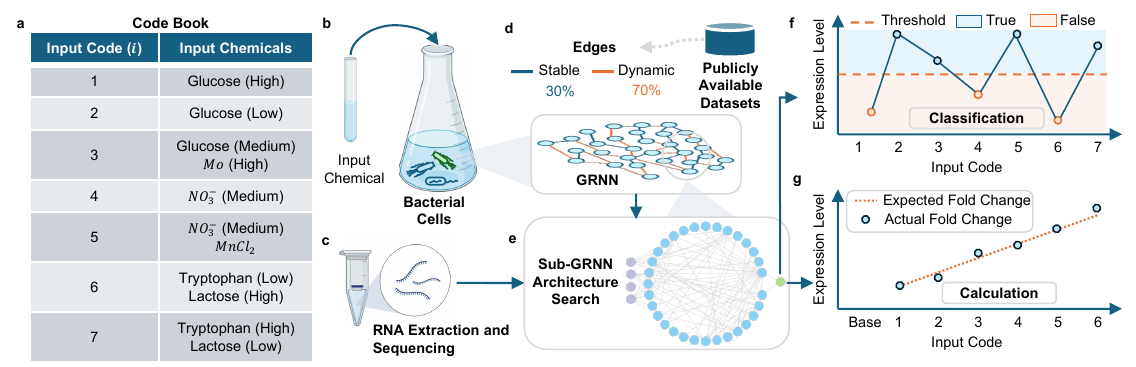} 
\vspace{-1.8em}
\caption{
\textbf{Overview of sub-GRNN architecture search and task-specific gene expression encoding.}
\textbf{a,} Codebook defining seven distinct input chemical combinations that correspond to integer number 1 - 7. 
\textbf{b,} \textit{E. coli} K-12 cells are exposed to the chemical inputs to elicit transcriptional responses.  
\textbf{c,} RNA is extracted and sequenced to obtain gene expression profiles. 
\textbf{d,} A GRNN is inferred from publicly available datasets, where we found that 30\% of gene-gene edges are stable and 70\% exhibit dynamic behavior across input conditions. 
\textbf{e,} A sub-GRNN is identified using a searching algorithm on the GRNN that also discover paths linking input-responsive genes to output genes whose expression aligns with the expected behavior. 
\textbf{f,} For classification tasks, gene expression levels are compared against a predefined threshold to assign bianry pattern. 
\textbf{g,} For calculation tasks, output genes are selected based on their fold-change correlation with expected numerical trends. Figure created with BioRender.com.
}
\vspace{-1em}
\label{fig:Concept_fig_V2}
\end{figure*}

This article introduces a framework (Figure~\ref{fig:Concept_fig_V2}) aimed at extracting sub-GRNN networks from full the GRNN via a search algorithm that identifies functional subnetworks solving a suite of mathematical problems, with the GRNN serving as a library of  computing structures defined by task-specific gene expression patterns. These sub-GRNNs resemble  artificial neural networks structures  found in AI \cite{SOMATHILAKA2023100118, RATWATTE2024100158, NEURIPS2024_1d35a777}. 
Our approach is based on assigning a integer value of \(i = 1, \dots, 7\), which we refer to as input codes, to a particular type of chemical condition that is applied to the bacteria.
As shown in Figure~\ref{fig:Concept_fig_V2}a, the input codes \(i = 1, \dots, 7\) represent combinations of glucose (C\(_6\)H\(_{12}\)O\(_6\)), nitrate (NO\(_3^-\)), tryptophan (C\(_{11}\)H\(_{12}\)N\(_2\)O\(_2\)), lactose (C\(_{12}\)H\(_{22}\)O\(_{11}\)), and trace metal MnCl\(_2\) concentrations.
These chemicals (Figure~\ref{fig:Concept_fig_V2}b) are added to \textit{E. coli} K-12 cultures, and RNA extraction and sequencing are performed at three time points (6, 30, and 70 hours) (Figure~\ref{fig:Concept_fig_V2}c).
\textcolor{red}{However, a step before this is constructing the full digital GRNN model. To do this,} we \textcolor{red}{develop }
the GRNN from publicly available datasets to  obtain a more comprehensive and reliable network, enabling to capture both stable and condition-specific regulatory gene-gene interaction edges (Figure~\ref{fig:Concept_fig_V2}d) \cite{Sonawane}. Identifying stable edges is essential because sub-GRNNs containing such edges can maintain consistent output expression across varying chemical inputs, supporting stable and reliable computing (See Supplementary Note~2 and Fig.~S2 in the Supplementary Information for a detailed explanation of the stable edges analysis).   
We then identify genes whose expression patterns match the solution of the task and trace their upstream regulatory pathways to extract the corresponding sub-GRNN (Figure~\ref{fig:Concept_fig_V2}e). 
The study considers both classification and calculation problems as summarized in Table~\ref{tab:problem-types}. 
For classification tasks, gene expression is binarized using a threshold rule (Figure~\ref{fig:Concept_fig_V2}f), while for calculation tasks, output genes are selected based on their fold-change alignment with mathematical trends (Figure~\ref{fig:Concept_fig_V2}g). 

\adrian{We then systematically characterize each subnetwork’s computing reliability and stability 
through three complementary analyses. In gene-wise perturbation, we perturb individual hidden-layer genes within the sub-GRNN to assess their influence on output expression and computing reliability \cite{baumstark2015propagation}. In collective perturbation, we evaluate performance under simultaneous multi-gene perturbations. We use the Lyapunov-based stability analysis to determine the critical perturbation thresholds beyond which computing becomes unstable \cite{1428924,LIU201613}.}
Together, these methods establish a 
biocomputing library of mathematical solvers embedded within living bacterial cells, harnessing native transcriptional dynamics for stable mathematical calculation tasks and classification.


\begin{table}[t!]
\caption{Mathematical Calculation and Classification Tasks with their Expected Outputs.}
\label{tab:problem-types}
\begin{tabular}{@{}p{6.8cm}p{3cm}p{7cm}@{}}
\toprule
\textbf{Problem} & \textbf{Problem Type} & \textbf{Expected Output  (for $i \in \{1,\dots,7\}$)} \\
\midrule
1) What is the \( i^{\text{th}} \) Fibonacci number? &  & \( 1, 1, 2, 3, 5, 8 \) \\
2) What is the Collatz Step Count for $i$ ? & Mathematical Calculation & \( 0, 1, 7, 2, 5, 8, 16 \) \\
3) What is \( i \) multiplied by \( m \), where \( i \in \{1,\dots,7\} \) and \( m \in \{2,3,4,5\} \) ?
 &  & \( m \times i \) \; (\(m=2,3,4,5\)) \\
\midrule
4) Is \( i \) a lucky number? & \multirow{4}{*}{Classification} & True for \( i = 1, 3, 7 \); false otherwise \\
5) Is \( i \) a prime number? &  & True for \( i = 2, 3, 5, 7 \); false otherwise \\
6) Is \( i \) a Fibonacci number? &  & True for \( i = 1, 2, 3, 5 \); false otherwise \\
7) Is the repeating cycle length of \(1/i\) equal to 1?
 &  & True for \( i = 3, 6 \); false otherwise \\
\bottomrule
\end{tabular}
\vspace{-1em}
\end{table}

\section*{RESULTS}


\adrian{In this section, we present sub-GRNNs identified for mathematical calculation and classification problems using a novel search algorithm we propose to match gene expression patterns with the expected output. For the calculation problems, we compute fold-changes relative to the baseline condition ($i = \textit{base}$) and search for genes whose patterns match expected values across all input codes ($i = 1$ to $6$) in both replicates for reproducibility. This is repeated at each timepoint, $t=6, 30, 70$ hours, corresponding to the transcriptomic data collected under each input condition, where the most suitable output gene is identified (Supplementary Algorithm~1). The variables and their definitions used in the search algorithms are summarized in Supplementary Table~1.
For classification problems, we search for genes whose expression patterns match a predefined binary pattern  across all input codes ($i=1,2,..., 7$) in both replicates, using the mean expression as a threshold (Supplementary Algorithm 2 and 3). 
We then select the best output gene by computing the distances from the threshold to the nearest expression values above and below it in each class. The midpoint of these distances defines the decision boundary, and the gene with the maximum  distance is selected. }

\adrian{We extract the sub-GRNN by tracing upstream connections from the output gene identified by the search algorithm to the chemically responsive input genes.} 
\adrian{
As shown in Table~\ref{tab:problem-types}, we analyze seven tasks, including the \(i^{\text{th}}\) Fibonacci number, the Collatz step count for \(i\), and multiplication as calculation problems.
The classification tasks are identifying Lucky numbers, Fibonacci numbers, prime numbers, and the repeating-cycle length in the decimal expansion of \(1/i\).
}

\adrian{To evaluate sub-GRNN stability and reliability of computing, we perform gene-wise perturbation by independently adding perturbation to each gene in the hidden layer of the sub-GRNN and measuring its effect on the output (Supplementary Algorithm~5) \cite{KATEBI2020101150}. Perturbation levels are drawn from a zero-mean Gaussian distribution (\ref{eq:state_eqn}, \ref{eq:noise_eqn}) scaled by the fold factor \(\alpha\), representing perturbation magnitude, and the noise variance \(\sigma^2\), representing perturbation spread.  
For calculation tasks, we compute the deviation in fold-change after the perturbation and calculate the coefficient of determination $R^2$ using ~\ref{eq:r_squared}. The mean $R^2$ across perturbation levels defines the error sensitivity of the gene 
(\ref{eq:mean_r_squared}). We also define the \emph{Criticality} of a gene 
as the ratio of its outward degree centrality to its mean \(R^2\) across perturbation levels (\ref{eq:criticality}), which captures both the network influence and perturbation sensitivity. Critical genes 
are those whose perturbation significantly alters the output-gene expression, where higher criticality indicates greater potential to disrupt computing of the sub-GRNN.
\textcolor{red}{Therefore, our perturbation analysis is taking the highest critical genes and perturbing them to determine its impact on the error at the output.} For the classification tasks, we compute the Hamming distance between the perturbed and unperturbed binary outputs across input codes (\ref{eq:hamming}), where the Hamming distance corresponds to the number of misclassifications. The classification-based criticality is defined as the product of outward degree centrality and the summed Hamming distance over all perturbation levels (\ref{eq:criticality_class}).
}

\adrian{To evaluate the collective impact of biologically plausible perturbations, we perform collective perturbation analysis (Supplementary Algorithm~6) \cite{KATEBI2020101150, otoupal2018multiplexed}. Unlike single gene 
perturbation, this method perturbs a set of genes \(\mathcal{P}\) simultaneously and models multi-gene disruptions. Perturbations are sampled independently for each \(p \in \mathcal{P}\) from a zero-mean Gaussian distribution scaled by fold factor \(\alpha\) and noise variance \(\sigma^2\) (\ref{eq:state_eqn}, \ref{eq:noise_eqn}), and propagated through correlation-weighted gene-gene interaction  edges \cite{baumstark2015propagation}. The resulting cumulative effect is added to the unperturbed output-gene expression. We evaluate sub-GRNN reliability under collective perturbation by computing coefficient of determination \cite{article} \(R^2\)  for calculation or Hamming distance  for classification as we sequentially perturb the \(k\) most critical genes from 
 gene-wise analysis, starting with the most critical gene and incrementally adding one gene 
at a time up to \(k = 10\).}

\adrian{Using Lyapunov-based stability \cite{LIU201613, 1428924}  analysis (see Methods section), we determine the maximum perturbation level a sub-GRNN can tolerate while maintaining stable and reliable computing. The critical point of instability is defined as the perturbation level \(s\) of the Lyapunov derivative \(\frac{d\mathcal{V}}{ds} = 0\) (\ref{eq:Lyapunov_derivative}), where \(\mathcal{V}\) denotes the Lyapunov function  that captures the effect of hidden-layer gene perturbations on output-gene expression and its impact on tolerable level of error $\epsilon$.
This marks the transition beyond which the computing error, defined as the deviation between unperturbed and perturbed output gene expression, rises sharply. Our analysis is to analytically determine the pertubation level before instability arises. We perform this analysis for both the most and least critical gene 
 and determine the cut-off perturbation level for each input code.
}



\subsection*{ What is the \( i^{\text{th}} \) Fibonacci number? }
%

\adrian{The first sub-GRNN discovered for the mathematical calculation task is to approximate the Fibonacci numbers, where each input code \(i = 1,\dots,6\) together with the base condition maps to the \(i^{\text{th}}\) Fibonacci number, defined as \(F(i) = F(i{-}1) + F(i{-}2)\). Using the search algorithm (Supplementary Algorithm~1), the gene \textit{b4056} is identified  at \(t = 70\) as the output. This is because the fold-change across the input codes matches the Fibonacci  number sequence in both replicates (Figure~\ref{fig:nth_FIBONACCI}a(I)), and the associated sub-GRNN is depicted in Figure~\ref{fig:nth_FIBONACCI}a(II). The increasing fold-changes of \textit{b4056} mirror the exponential growth of the Fibonacci number sequence, indicating that this gene follows the task's exponential pattern as input chemicals \(i = 1,\dots,6\) are introduced sequentially.}
Gene-wise pertubation analysis shows that 
(Figure~\ref{fig:nth_FIBONACCI}a(III)), a number of genes, specifically gene \textit{b3067}, exhibit high criticality. 
This follows from the criticality \textcolor{red}{calculation}, where high centrality combined with low  coefficient of determination \(R^2\)  amplifies criticality (\ref{eq:criticality}). 
Certain genes can have high centrality but yields minimal output error when perturbed due to the high coefficient of determination \(R^2\) such as gene \textit{b1988}.
The near-exponential growth of the Fibonacci number sequence makes higher input codes more sensitive to perturbation, so perturbing the most critical gene \textit{b3067} on the regulatory path to the output gene \textit{b2927} amplifies deviations in the output expression.
\adrian{Therefore, to evaluate how perturbations propagate through the Fibonacci sub-GRNN, we analyze the output error under increasing perturbation levels applied to the most critical gene, \textit{b3067}. 
Lower input codes (\(i = 1\)) resulted in stable near-zero error, as the gene expression correlation matrix \(\mathbf{W}\) in \ref{eq:state_eqn} attenuates the perturbation  \(\mathbf{u}(t; \alpha, \sigma)\), 
while stabilizing the output expression under the metabolically stable Glucose (Low) condition~\cite{zaslaver2004just}.
In contrast, for other input codes (\(i = 2, 3, 4, 5, 6\)), the 
perturbations from \textit{b3067} affects 
the output more strongly. 
This is due to the stress-responsive conditions corresponding to \(i = 2, 3, 4, 5, 6\) increasing  gene expression correlations~\cite{zaslaver2004just}, \cite{lopez2008tuning}  that results in larger deviations in output expression of the Fibonacci sub-GRNN.
}

According to the collective perturbation analysis (Figure~\ref{fig:nth_FIBONACCI}a(V)), the mean coefficient of determination \(R^2\) decreases steadily with the number of perturbed genes \(k\).
This suggests that output reliability is primarily governed by a small set of highly critical genes, as the loss of \(R^2\) is sharp when perturbing the top six genes and more gradual beyond that point.
Higher perturbation levels further reduce \(R^2\) across most input codes, confirming that noise impairs reliable computing even though some inputs remain less sensitive.
\adrian{
Lyapunov based stability analysis shows that when the most critical gene \textit{b3067} is perturbed, the sub-GRNN remains stable up to perturbation level \(\alpha = 14.12\), \(\sigma^2 = 1.50\) for input code \(i = 1\), as output expression remains largely unchanged, consistent with the near-zero error observed in Figure~\ref{fig:nth_FIBONACCI}a(IV). In contrast, input codes \(i = 2\), \(3\), and \(6\) exhibit instability at lower perturbation levels (\(\alpha = 5.88\), \(5.40\), and \(5.01\)), since even small perturbations cause abrupt changes in output expression, indicating reduced tolerance to noise under those input conditions (Figure~\ref{fig:nth_FIBONACCI}a(VI)). 
The observed effect reflects input-specific modulation of  gene expression correlation matrix, which shapes how perturbations propagate through the sub-GRNN and impact the exponential growth of Fiboancci sequence. For Glucose (Low) (\(i = 1\)), weaker correlations in \(\mathbf{W}\) (\ref{eq:state_eqn}) and smaller expression deviations \(\delta x_q(I, t)\) result in lower Lyapunov values \(\mathcal{V}(I, t; \alpha, \sigma)\) (\ref{eq:Lyapunov_fn}, \ref{eq:weight_lyapunov}), allowing higher perturbation tolerance. In contrast, stress-responsive inputs \cite{zaslaver2004just}, \cite{lopez2008tuning} such as NO\(_3^-\), Tryptophan, and Mo (\(i = 2, 3, 4, 5, 6\)) increase \(\delta x_q\) raising \(\mathcal{V}\) and triggering instability at lower perturbation levels.
The minimum threshold among all input codes, \(\alpha = 5.01\), defines the conservative upper bound for allowable perturbations that preserve computing stability.}
See Supplementary Note~5.1 and Figure~S3a for supporting results of the gene-wise perturbation and Lyapunov stability analysis for the \(i^{\text{th}}\) Fibonacci sub-GRNN.

\subsection*{What is the Collatz Step Count for $i$ ?}

\begin{figure*}[t!] 
\centering
\includegraphics[width=0.9\linewidth]{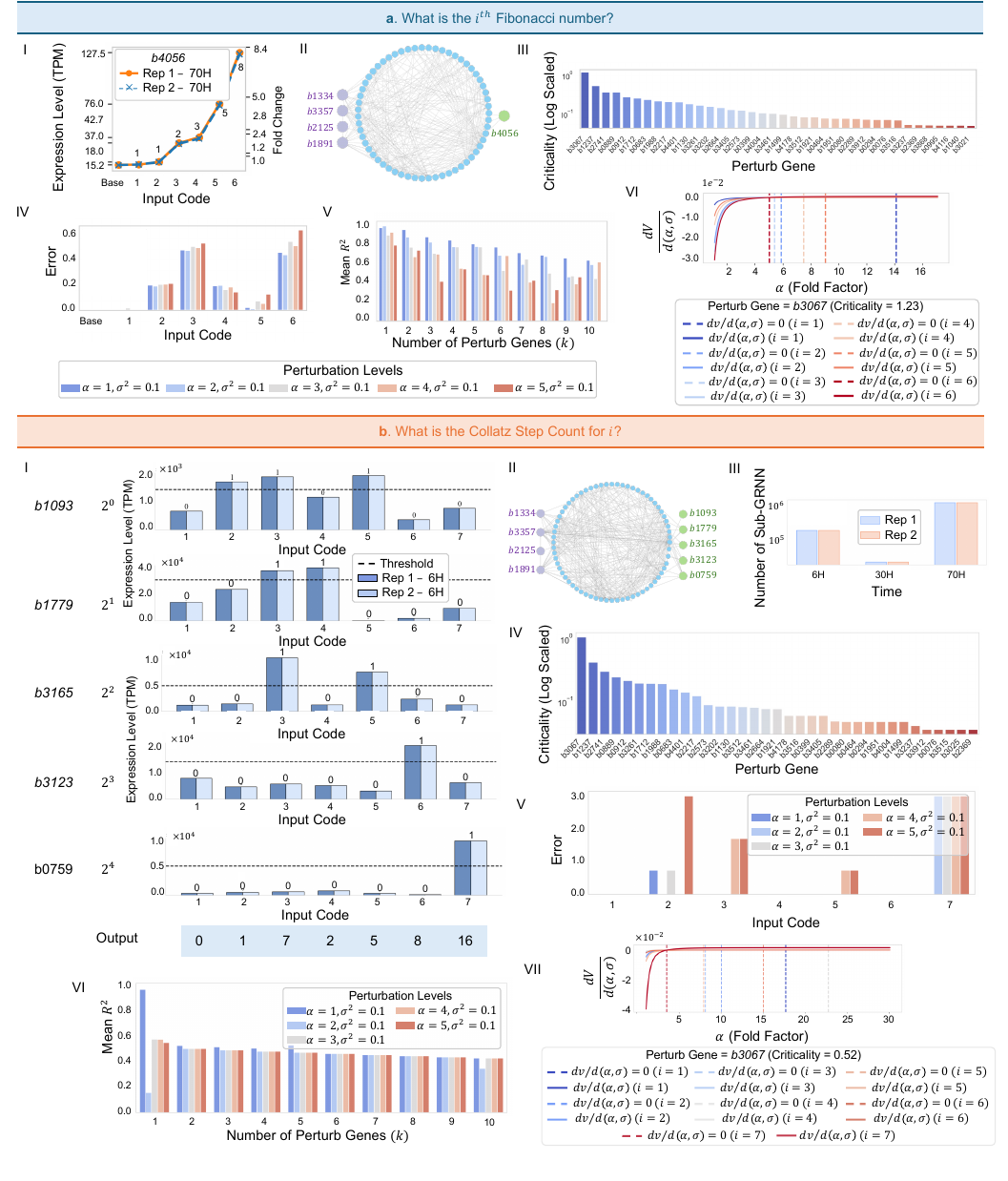} 
\vspace{-2em}
\caption{
\textbf{GRNN-based computing of the $i^{\text{th}}$ Fibonacci number (a) and Collatz step count (b).} 
\textbf{a,} Fibonacci number calculation task using GRNN. \textbf{I,} Output gene \textit{b4056} expression across input codes matching the pattern of Fibonacci sequence. \textbf{II,} Searched sub-GRNN with input-responsive genes in purple and intermediate genes in blue. \textbf{III,} gene-wise criticality scores. \textbf{IV,} Error under perturbation for each input code. \textbf{V,} Collective perturbation effects on mean \(R^2\). \textbf{VI,} Lyapunov stability analysis for most critical gene \textit{b3067}. 
\textbf{b,} Collatz step-count task using binary encoding. \textbf{I,} Expression of five output genes corresponding to binary weights \(2^0\)–\(2^4\). \textbf{II,} Sub-GRNN structure. \textbf{III,} Number of valid subnetworks over time for both experimental replicates. \textbf{IV,} Criticality score for each pertub gene. \textbf{V,} Input-specific error under gene-wise perturbation. \textbf{VI,} Collective perturbation impact on mean \(R^2\). \textbf{VII,} Lyapunov stability analysis for the most critical gene.
}
\vspace{-1.3em}
\label{fig:nth_FIBONACCI}
\end{figure*}

\adrian{ The Collatz step-count for $i$ represents the number of iterations needed to reach 1 by repeatedly applying the rule that if \(i\) is even it is divided by 2, and if it is odd then it is replaced with \(3i + 1\). Using a distinct binary pattern search (Supplementary Algorithm~4), we extract the sub-GRNN whose output gene expression pattern matches the Collatz step-count for \(i = 1, \dots, 7\).}
\adrian{
We search for genes whose expression patterns follow predefined binary pattern across input codes ($i = 1 \dots 7$) in both replicates. At each timepoint, a threshold is defined as the midpoint of the largest adjacent expression gap. The best-matching gene is selected by maximizing the distance between the closest expression values above and below the threshold (see Supplementary Algorithm~4). 
The search algorithm identifies five output genes including \textit{b1093}, \textit{b1779}, \textit{b3165}, \textit{b3123}, and \textit{b0759}, whose thresholded expression patterns at \(t = 6\) hours accurately reconstruct the Collatz step count across input codes using binary weights \(2^0\) to \(2^4\) (Figure~\ref{fig:nth_FIBONACCI}b(I)).
}
For example, input code \(i = 2\) yields an output of 1, and only \textit{b1093} (i.e., \(2^0\)) is expressed above threshold, with all others remaining below. At \(i = 3\), the output is 7, and genes \textit{b1093}, \textit{b1779}, and \textit{b3165} are all strongly activated, matching the binary pattern \(111\). 
\adrian{This pattern is driven by the input-specific regulation that enables binary-like gene expression~\cite{kaern2005stochasticity}, \cite{wong2015layering}, where consistent thresholded activation of output genes across replicates aligns with their respective binary weights (\(2^0\) to \(2^4\)), allowing precise reconstruction of the Collatz sequence \([0, 1, 7, 2, 5, 8, 16]\) through binary encoding of gene expression. To identify the underlying computing structure, we extract the sub-GRNN for Collatz step count task (Figure~\ref{fig:nth_FIBONACCI}b(II)).
}



To evaluate how many matching sub-GRNNs exist for the Collatz step-count task, we enumerate all sub-GRNNs whose output activation patterns match the expected binary outputs (Supplementary Algorithm~3). Figure~\ref{fig:nth_FIBONACCI}b(III) shows the number of such valid subnetworks at three time points. The  search algorithm used in this task yields over \(10^6\) matched sub-GRNNs at $t=70$ hours, substantially more than other tasks, highlighting how binary encoding expands the sub-GRNN search space.
\adrian{To pinpoint the most influential regulators in the Collatz computing, we conduct gene-wise perturbation analysis. A few genes, most notably \textit{b3067}, \textit{b2731}, and \textit{b4099} dominate, while the majority have minimal influence on the output (Figure~\ref{fig:nth_FIBONACCI}b(IV)).
 This outcome results from a sparse set of hidden-layer genes that disproportionately govern the binary, thresholded output pattern; consistent with the criticality (\ref{eq:criticality}), where high centrality and low \(R^2\) indicate strong, non-redundant impact on computing.}

 The output error under increasing perturbation of the most critical gene \textit{b3067} (Figure~\ref{fig:nth_FIBONACCI}b(V)) indicates that for lower input codes \(i=1, 4, 6\) remain relatively stable, while higher codes, particularly \(i=2, 3,\) and \(7\) exhibit substantial increases in error, with \(i=7\) reaching the maximum deviation across all perturbation levels.
 \adrian{This behavior is driven by input-dependent output activation and perturbation propagation from the critical gene \textit{b3067}. For input codes \(i=1, 4, 6\), output genes \textit{b1093}, \textit{b3165}, and \textit{b0759} remain inactive (below threshold) (Figure~\ref{fig:nth_FIBONACCI}b(I)), allowing the correlation matrix \(\mathbf{W}\) (\ref{eq:state_eqn}) to dampen perturbations \(\mathbf{u}(t; \alpha, \sigma)\). For higher codes (\(i=2, 3, 5, 7\)), \(\mathbf{W}\) propagates perturbations more strongly, leading these output genes to flip between inactive and active states and increasing output error (Figure~\ref{fig:nth_FIBONACCI}b(V)).
}

In the collective perturbation analysis (Figure~\ref{fig:nth_FIBONACCI}b(VI)), perturbing only the most critical gene (\(k=1\)) causes the mean coefficient of determination (\(R^2\)) to vary markedly with perturbation level \(\alpha\), highlighting its strong individual influence. However, as more genes are perturbed, mean \(R^2\) stabilizes around 0.4–0.5, regardless of perturbation strength. The steep initial drop from \(k=1\) to \(k=2\) reflects redundancy among the most critical genes, while subsequent additions cause smaller incremental declines.
This is because the Collatz step-count encoding depends on binary activation of output genes, where the most critical hidden-layer genes control multiple binary bits of the pattern. Many of the remaining genes have similarly low criticality (Figure~\ref{fig:nth_FIBONACCI}b(IV)), and their overlapping influence on the binary output encoding means that perturbing them together adds minimal disruption, leading to saturation \cite{hiratani2022stability}.

The Lyapunov-based analysis of the most critical gene \textit{b3067} (Figure~\ref{fig:nth_FIBONACCI}b(VII)) shows that 
 inputs \(i = 1\) and \(i = 4\) exhibit the highest tolerance, with stability maintained up to perturbation level \(\alpha = 17.72\), \(\sigma^2 = 1.86\) and \(\alpha = 22.78\), \(\sigma^2 = 2.37\), respectively. Inputs \(i = 2\) and \(i = 6\) become unstable at moderate thresholds (\(\alpha = 10.05\), \(\sigma^2 = 1.09\) and \(\alpha = 15.05\), \(\sigma^2 = 1.59\)). Inputs \(i = 3\), \(i = 5\), and \(i = 7\) reach criticality at lower perturbation levels with \(\alpha = 8.13\), \(\sigma^2 = 0.90\); \(\alpha = 7.96\), \(\sigma^2 = 0.89\); and \(\alpha = 3.54\), \(\sigma^2 = 0.44\), respectively. 
  \adrian{A key factor is the input-specific activation combined with perturbation propagation from the gene \textit{b3067} across the sub-GRNN.  
 For inputs \(i = 1, 2, 4, 6\),  output genes (\textit{b1093}, \textit{b3165}, \textit{b0759}) stay inactive (Figure~\ref{fig:nth_FIBONACCI}b(I)), leading to smaller expression changes \(\delta x_q(I, t)\), weaker correlations in \(\mathbf{W}\), and lower Lyapunov values \(\mathcal{V}(I, t; \alpha, \sigma)\) (\ref{eq:Lyapunov_fn},~\eqref{eq:weight_lyapunov}), which allows higher tolerance to perturbation. Inputs \(i = 3, 5, 7\) activate more output genes above threshold, increasing \(\delta x_q\) and allowing perturbations to propagate more strongly through \(\mathbf{W}\), causing instability at lower \(\alpha\).}
 The minimum threshold among all input codes, \(\alpha = 3.54\) at \(i = 7\), defines the conservative upper bound for allowable perturbations that preserve computing stability in the Collatz sub-GRNN. 
These results are further supported by additional analyses, including stable edge analysis, mean coefficient of determination \(R^2\) from gene-wise perturbations, and Lyapunov stability analysis of the least critical gene, presented in Supplementary Figure~S3b.

\subsection*{What is \( i \) multiplied by \( m \), where \( i \in \{1,\dots,7\} \) and \( m \in \{2,3,4,5\}? \)}

\begin{figure*}[t!] 
\centering
\includegraphics[width=\linewidth]{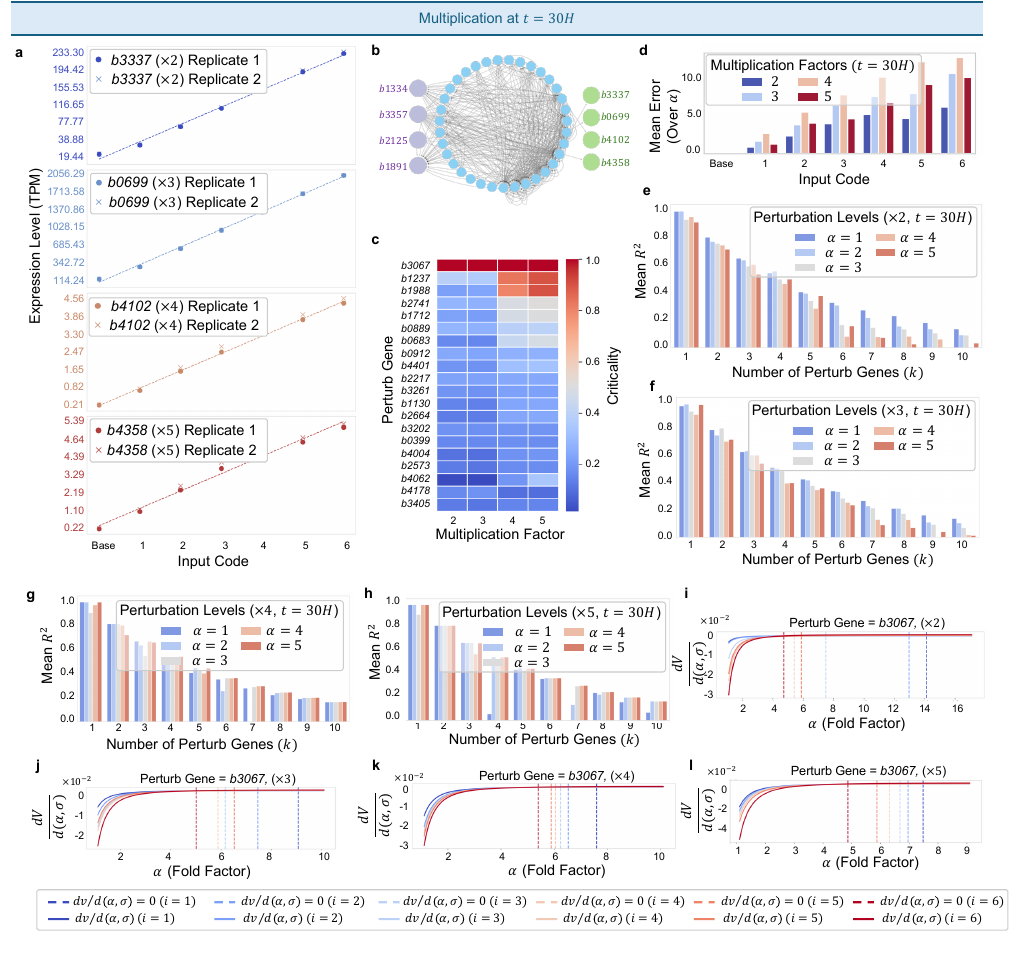} 
\vspace{-3.4em}
\caption{
\textbf{GRNN-based  multiplication at \textbf{t=30H}.} 
\textbf{a,} Expression profiles of output genes \textit{b3337}, \textit{b0699}, \textit{b4102}, and \textit{b4358}, corresponding to multiplication factors \(  2 \) to \(  5 \), respectively, across input codes. 
\textbf{b,} Extracted sub-GRNN showing input-responsive genes in purple, intermediate genes in blue, and output genes in green. 
\textbf{c,} Gene-wise criticality scores for each multiplication factor. 
\textbf{d,} Input-specific mean error under gene-wise perturbation for each multiplication factor. 
\textbf{e–h,} Collective perturbation impact on mean \(R^2\) for multiplication factors \( 2\) to \( 5\) under perturbation levels  \(\alpha=1\) to $5$ with fixed noise variance \(\sigma^2=0.1\). 
\textbf{i–l,} Lyapunov-based stability analysis for the most critical gene for each multiplication factor.
}
\vspace{-1.2em}
\label{fig:multiplication_main}
\end{figure*}

The multiplication task computes the products of input codes \( i \in \{1,\dots,7\} \) with fixed factors \( m \in \{2,3,4,5\} \).  
 \adrian{We identify genes whose expression scales proportionally with the input code for fixed multiplication factors using the search algorithm (Supplementary Algorithm~1). \textcolor{red}{The search algorithm identified} four output genes including \textit{b3337}, \textit{b0699}, \textit{b4102}, \textit{b4358}, at \(t = 6, 30\) hours that gave expression patterns that match the multiplication factors 2, 3, 4, and 5, respectively. Since the closest matches in output-gene expression levels occur at \(t = 30\) hours, we describe the key results in this section based on that time point, with supportive results provided in Supplementary Figure~S6.
 \textcolor{red}{At \(t = 70\) the search algorithm also identified} genes \textit{b3337} and \textit{b0699} for multiplication factors 2 and 3 (see Supplementary Figure~S7 for the expression profiles of output genes and the number of sub-GRNNs identified at \(t = 6\) and $70$ hours.).  Figure~\ref{fig:multiplication_main}a shows the expression trajectories of the output genes at \(t = 30\) hours across input codes.} \textcolor{red}{The plot shows the} expression increases linearly from the base, maintaining an approximate fold-matched progression across input codes and both replicates, consistent with multiplicative operation. 
For instance, \textit{b4358} shows a near fivefold change from base to input 1 (\(1.1 / 0.22 \approx 5\)), and \textit{b0699} approximates a threefold increase (\(342.72 / 114.24 \approx 3\)). 
This linear pattern is shaped by hidden-layer gene-to-gene interactions that amplify and propagate input expression levels through the sub-GRNN, producing an approximately linear output response.
\adrian{Figure~\ref{fig:multiplication_main}b shows the sub-GRNN for this multiplication task, where green nodes (\textit{b3337}, \textit{b0699}, \textit{b4102}, \textit{b4358}) represent output genes encoding fold factors 2 to 5, blue nodes form the hidden layer, and \textcolor{red}{the} purple nodes correspond to input-responsive genes.
}

Figure~\ref{fig:multiplication_main}c displays the criticality of perturbation genes across multiplication factors 2–5. 
\rev{Gene \textit{b3067} consistently ranks highest in criticality across all multiplication factors because its perturbation strongly propagates through the amplification pathways required for scaling output expression, making it essential for maintaining the output proportional to the multiplication factor.} 
Gene \textit{b1237} also shows high criticality, particularly for multiplication factors 4 and 5, suggesting these outputs rely on shared upstream pathways. As the multiplication factor increases, the number of moderately critical genes expands (e.g., \textit{b1988}, \textit{b2741}). This pattern arises because higher multiplication factors need greater expression amplification, 
which depends more on upstream genes such as \textit{b1237} and \textit{b1988} that lie on the paths to output genes for \textcolor{red}{multiplication} factors \textcolor{red}{of} 4 and 5. 
\rev{Perturbing hub genes such as \textit{b3067, b1237, b1988}, which have high degree centrality and connect indirectly to all four output genes, propagates perturbation broadly through the network and changes 
the expected linear scaling  of output expression under higher multiplication factors \cite{barabasi2004network}, \cite{marbach2016tissue}.} 
Supplementary Figure~S8 illustrates the criticality distribution under gene-wise perturbation at \(t = 6\) and $70$ hours.

\adrian{We examine the error for each input code under perturbation of the most critical gene \textit{b3067}, where the error is averaged over all tested perturbation levels \(\alpha=1,2,...,5\) with fixed noise variance \(\sigma^2 = 0.1\) (Figure~\ref{fig:multiplication_main}d). \textcolor{red}{The} error increases with input code for all multiplication factors, showing that genes expected to have higher expression levels are more affected by perturbations, especially at input codes 5 and 6. \rev{In contrast, at multiplication factor \(2\), the error remains consistently low across input codes.}
 \rev{This behavior arises because higher input codes require proportionally larger expression changes to preserve the linear scaling required for multiplication. \textcolor{red}{However, the} perturbations to critical genes change these slopes, breaking the expected linear \textcolor{red}{relationship} between \textcolor{red}{the} input code and output-gene expression.} Lower multiplication factors such as 2 follow a gentler linear trend with smaller gene expression changes.}



In the collective perturbation analysis for multiplication factor \(2\) (Figure~\ref{fig:multiplication_main}e), the mean coefficient of determination \(R^2\)  drops quickly as more critical genes ($k$) are perturbed simultaneously, especially up to \(k = 5\), and levels off near 0.2 after \(k = 6\) (Figure~\ref{fig:multiplication_main}e). 
Figure~\ref{fig:multiplication_main}f shows the collective perturbation analysis for factor \( 3\). While the overall trend is similar to multiplication factor \(\ 2\), the drop in \(R^2\) is more gradual, with moderate \(R^2\) values still observed up to \(k = 5\). Higher perturbation levels (\(\alpha = 4, 5\)) further accelerate the decline in \(R^2\).
Figure~\ref{fig:multiplication_main}g,h show collective perturbation effects for the multiplication factor \(\ 4\) and \( 5\). In both cases, the decline in \(R^2\) is slower compared with factors \(2\) and \(3\). For multiplication factor \(4\), \(R^2\) drops below 0.5 by \(k = 5\) (Figure~\ref{fig:multiplication_main}g). For factor \(5\), the decline is more gradual and levels off around 0.2 even as more genes are perturbed (Figure~\ref{fig:multiplication_main}h).  
This behavior can be explained by how perturbations propagate to the output genes. For higher multiplication factors such as 4 and 5, propagation is weaker because output-gene expression relies on a broader set of critical genes, whereas for factors 2 and 3 the output depends on fewer genes and is, therefore, more sensitive to perturbation.  


\rev{Lyapunov-based stability analysis of the most critical gene \textit{b3067} for multiplication factor \(2\) (Figure~\ref{fig:multiplication_main}i) reveals that instability emerges gradually across input codes.} Lower input codes \(i = 1, 2\) remain stable up to larger perturbation level \(\alpha > 12\), while higher input codes \(i = 5, 6\) transition earlier (\(\alpha = 5.86, 4.72\)).
Under multiplication factor \( 3\) (Figure~\ref{fig:multiplication_main}j), the transition thresholds narrow considerably, ranging from \(\alpha = 5.02\) to \(9.06\). 
\rev{Multiplication factor \( 4\) (Figure~\ref{fig:multiplication_main}k) demonstrates increased sensitivity to perturbation  occurring within the range \(\alpha = 5.36\)–\(7.57\), with input code \(i = 6\) consistently destabilizing first.} 
For multiplication factor \( 5\) (Figure~\ref{fig:multiplication_main}l), the earliest instability is observed (\(\alpha = 4.83\) for \(i = 6\)), and all other input codes fall below \textcolor{red}{the} threshold within a narrow window (\(\alpha = 4.83\) to \(7.46\)). 
Comparing across multiplication factors, input codes \(i = 5, 6\) consistently exhibit the earliest transitions.
Input codes \(i = 1, 2\) remain the most tolerant across all factors, with delayed thresholds exceeding \(\alpha = 12\) under multiplication factor \( 2\).
When perturbing the most critical gene in the multiplication sub-GRNN, disproportionately disrupts output-gene expression in the input conditions of \(i = 5,6\).
This disruption amplifies the deviations \(\delta x_q(I,t)\), increases the Lyapunov function \(\mathcal{V}(I,t;\alpha,\sigma)\) (Eqs.~\ref{eq:Lyapunov_fn},~\ref{eq:weight_lyapunov}), and triggers earlier instability.
The minimum stability threshold across all input–factor pairs occurs at \(\alpha = 4.72\) for input \(i = 6\) and multiplication factor \( 2\), defining the most conservative upper bound on tolerable perturbation for stable computing. 

\subsection*{Is \(i\) a Lucky Number?}

\begin{figure*}[t!] 
\centering
\includegraphics[width=\linewidth]{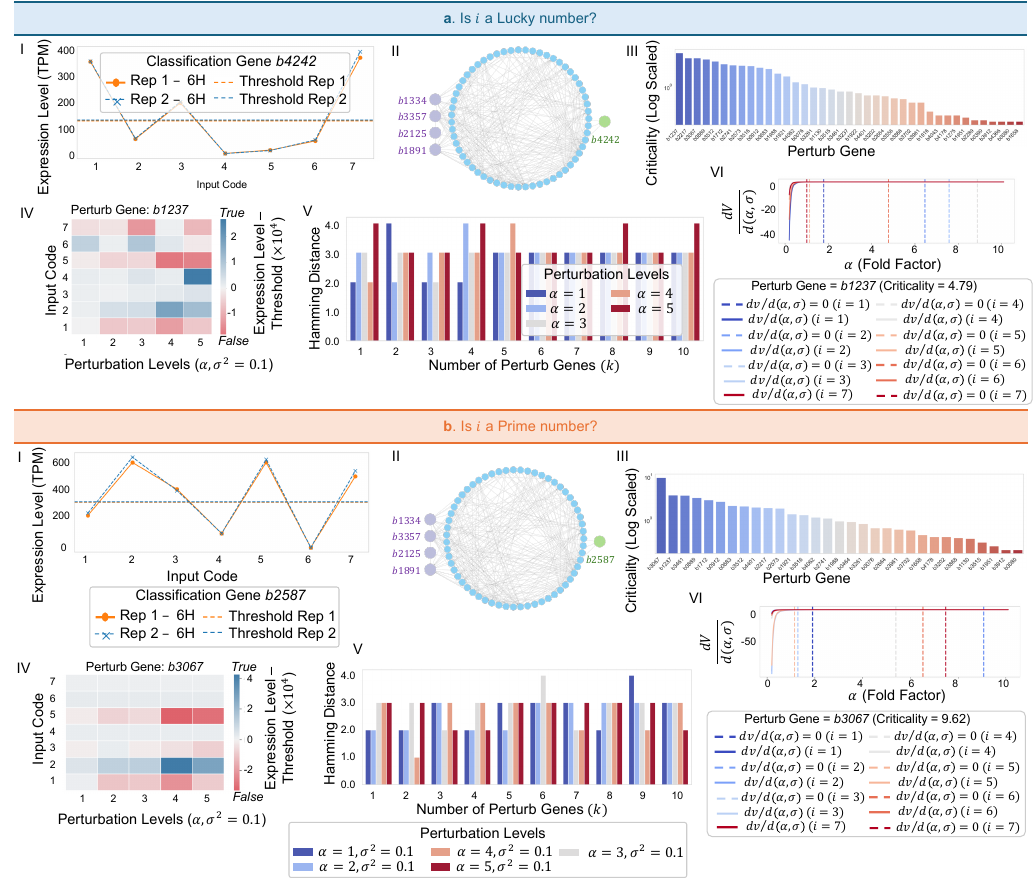} 
\vspace{-2em}
\caption{
\textbf{GRNN-based Lucky number classification (a) and prime number classification (b).} 
\textbf{a,} Classification of whether input \(i\) is a Lucky number using output-gene \textit{b4242}. 
\textbf{b,} Classification of whether input \(i\) is a prime number using output-gene \textit{b2587}. 
\textbf{I,} Expression profile of the output gene across input codes with decision thresholds for each replicate. 
\textbf{II,} Extracted sub-GRNN structure connecting input-responsive genes (purple), intermediate regulators (blue), and the classifier output genes (green). 
\textbf{III,} The criticality scores under gene-wise perturbation. 
\textbf{IV,} Expression deviation from threshold under perturbation of most critical gene (\textit{b1237} in a, \textit{b3067} in b). 
\textbf{V,} Collective perturbation impact on classification accuracy measured by Hamming distance across perturbation levels \(\alpha=1\) to $5$ with fixed noise variance \(\sigma^2=0.1\). 
\textbf{VI,} Lyapunov-based analysis for the most critical gene.
}
\vspace{-1em}
\label{fig:lucky}
\end{figure*}

The first sub-GRNN identified for the classification task determines whether an input code \(i \in \{1, \dots, 7\}\) is a Lucky number, which is an integer that survives a recursive sieve process similar to the \emph{Sieve of Eratosthenes} \cite{10.1145/321371.321373}. The expected Lucky numbers in this range are \(1, 3, 7\). Using the search algorithm for classification tasks  (Supplementary Algorithms~2 and 3), we identified \textcolor{red}{a sub-GRNN with} output-gene \textit{b4242}, whose expression at \(t = 6\) hours matches the expected binary pattern (Figure~\ref{fig:lucky}a(I)). The input codes \(i = 1, 3, 7\) exceed the classification threshold in both replicates, while all the others remain below. \rev{This behavior arises due to \textcolor{red}{the chemical compounds of input codes} \(i = 1, 3, 7\), which correspond to Lucky numbers, activate gene-to-gene interactions in the sub-GRNN that drive high expression of \textit{b4242}.} We extract the sub-GRNN surrounding the output gene \textit{b4242}, which is presented in  Figure~\ref{fig:lucky}a(II).

\rev{The gene-wise perturbation analysis  reveals a long-tailed criticality distribution, indicating that more than 10 perturbed genes have criticality (\(>1\)) and   exert substantial influence on classification reliability and stability (Figure~\ref{fig:lucky}a(III)). 
} 
 The impact of perturbing the most critical gene \textit{b1237} is shown in Figure~\ref{fig:lucky}a(IV), where   each cell   represents the difference between the expression level of classifier gene \textit{b4242} and its decision threshold at \(t = 6\) hours.
Positive values (blue) indicate correct classification, while negative values (red) denote misclassifications. The color intensity reflects the distance between the expression level and the decision threshold. 

For input codes \(i=1,5,7\), perturbing the gene \textit{b1237} alters the output gene expression, leading to misclassification once perturbations exceed \(\alpha > 1\), whereas \(i=6\) remains correctly classified at low to moderate perturbations and misclassifies only beyond \(\alpha > 4\).  
For \(i=2,3,4\), the outputs remain correctly classified across perturbation levels \(\alpha=1\!-\!5\), although the expression levels move closer to the threshold while still preserving correct classification.
\rev{This behavior arises because perturbing the most critical gene \textit{b1237}, which connect indirectly to the output gene \textit{b4242}, alters expression levels and propagates perturbations strongly  under the input conditions \(i=1,5,7\).} 

In the Lucky number sub-GRNN, collective perturbations yield mostly moderate Hamming distances (2–3) across number of pertub \(k\) and pertubation level \(\alpha\).  The Hamming distance reaches its maximum value of four when \(k = 1,4,5,8,10\) genes are perturbed (Figure~\ref{fig:lucky}a(V)).
\rev{This behavior occurs because the Lucky-number sub-GRNN relies on a high-criticality subset of hidden-layer genes that regulate the output gene \textit{b4242}. When these genes are perturbed collectively, the gene expression pattern for some input codes (\(i=1,5, 7\)) is disrupted across the pertubaiton levels, leading to misclassification.} 
Lyapunov stability analysis of the most critical gene \textit{b1237} (Figure~\ref{fig:lucky}a(VI)) shows that the sub-GRNN becomes unstable in input conditions \(i=5,7,1,6\) at low perturbation thresholds \(\alpha=1.2, 1.9, 1.95,\) and \(4.5\), respectively. 
For input codes \(i=2,3,4\), the sub-GRNN tolerate higher perturbation levels with thresholds at \(\alpha=6.47, 7.58,\) and \(8.89\), respectively. This behavior arises because small perturbations propagate strongly through the sub-GRNN in input conditions \(i=1,5,7\), altering the expression of \textit{b4242} below the threshold for \(i=1,7\) and above the threshold for \(i=5\).  
These changes sharply increase the Lyapunov function \(\mathcal{V}(\alpha, \sigma)\) (\ref{eq:classify_lyapunov}) and lead to misclassifications.
 \rev{\textcolor{red}{Therefore}, the Lucky number classification sub-GRNN can tolerate perturbations up to \(\alpha = 1.9\) and \(\sigma^2 = 0.13\) across input conditions, while still maintaining stable and reliable classification.}
Additional results of stable edge analysis and coefficient of determination \(R^2\) distributions under  gene-wise perturbation for the Lucky number classification  sub-GRNN are presented in Supplementary Figure~S4a.

\subsection*{Is $i$ a Prime Number?}
Using the search algorithm for classification tasks (Supplementary Algorithms~2 and 3), we search for a gene whose expression pattern matches the binary classification of prime numbers. Input codes \(i = 2, 3, 5, 7\) are labeled as prime numbers divisible only by 1 and itself, and \(i = 1, 4, 6\) are labeled as non-prime numbers.  
We identified output gene \textit{b2587} as the classifier for the prime number task, with expression at \( t = 6 \) hours matching the expected pattern (Figure~\ref{fig:lucky}b(I)), and the corresponding sub-GRNN is shown in Figure~\ref{fig:lucky}b(II).    


\rev{The gene-wise perturbation analysis of the prime number sub-GRNN (Figure~\ref{fig:lucky}b(III)) shows a gradual decline in criticality distribution.
The gene \textit{b3067} shows the highest criticality, followed by a sharp drop to \textit{b3461} and \textit{b1237}, after which most other hidden-layer genes gradually decrease to low values (\(<5\)).}  
The impact of perturbing most critical gene \textit{b3067}  is summarized in Figure~\ref{fig:lucky}b(IV).
As the perturbation level increases from \(\alpha=1\) to \(5\), the output gene expression falls below the decision threshold in input codes \(i=3,5\), resulting in false negatives. For the non-prime number input code \(i=1\), the expression instead rises above the threshold, producing false positives. 
In input codes \(i=2,4,6,7\), the output-gene expression matches the expected classification pattern across perturbations \(\alpha=1\!-\!5\). 
This behavior arises because input conditions corresponding to input codes \(i = 3, 5\) activate regulatory pathways that drive \textit{b2587} expression near the decision threshold, making them sensitive to small perturbations that repress expression and cause false negatives. 
The perturbations to the most critical gene do not strongly propagate in input conditions \(i=2,4,6,7\), allowing the output to remain correctly classified but closer to the threshold across \(\alpha=1\!-\!5\).
Additional results are provided in Supplementary Note~5.4 and Supplementary Figure~S4b, which include results on mean coefficient of determination \(R^2\) from gene-wise perturbation and Lyapunov stability analysis of the least critical gene, which shows much higher perturbation tolerance than perturbing the most critical gene.

In the collective perturbation analysis (Figure~\ref{fig:lucky}b(V)), even perturbing the most critical gene (\(k=1\)) produces a nonzero Hamming distance in the range of 2-3. Beyond \(k=1\), perturbing more genes does not markedly increase the Hamming distance, and it remains near \(2\text{–}3\) for most \(k\) across perturbation levels \(\alpha\).
Exceptions occur at \(k=6,9\), where some perturbation levels yield the maximum Hamming distance of 4.
\adrian{This behavior arises because the most critical govern the key regulatory paths with high degree centrality that lead to the output gene  \textit{b2587}, enabling strong perturbation propagation and significant expression changes that drive  misclassifications. Perturbing additional genes beyond the most critical gene yields only marginal impact on the output, because criticality drops steeply after the most critical gene \textit{b3067} (Figure~\ref{fig:lucky}b(III)).}
Figure~\ref{fig:lucky}b(VI) presents the Lyapunov-based stability analysis under  perturbation of the most critical gene \textit{b3067}. 
The sub-GRNN becomes unstable in input conditions \(i=1,3,5\) at perturbation thresholds \(\alpha=1.91, 0.95,\) and \(0.9\), indicating low tolerance to perturbation. 
For inputs \(i=2,4,6,7\), the sub-GRNN tolerates higher perturbations with thresholds at \(\alpha=9.07, 5.40, 6.53,\) and \(7.48\), respectively.
\adrian{This behavior arises because, for inputs \(i = 2,4,6,7\), the expression of the output gene \textit{b2587} lies far from the classification threshold under unperturbed conditions, making it less sensitive to perturbation. As a result,  large changes to the upstream critical gene \textit{b3067} do not drive \textit{b2587} across the decision boundary, keeping \(\mathcal{V}(\alpha, \sigma)\) (\ref{eq:classify_lyapunov}) low and preserving classification stability.}
Therefore, the earliest instability at \(\alpha = 0.9\), \(\sigma = 0.22\) across input codes defines the maximum perturbation tolerance for reliable and stable classification of the sub-GRNN.  

\subsection*{Is $i$ a Fibonacci Number}
Unlike the $i^{th}$ Fibonacci number task that computes the Fibonacci numbers, this task classifies whether an input code \( i \) belongs to the set of Fibonacci numbers.  
\rev{Using the search algorithm for classification tasks (Supplementary Algorithms~2 and 3), we identify an output-gene whose expression pattern matches the  classification pattern of Fibonacci numbers. The input codes \( i = 1, 2, 3, 5 \) are labeled as Fibonacci numbers, since each is obtained by adding the two preceding numbers.}
We identified gene \textit{b2927} as the output-gene for this task, with expression at \(t = 6\) 
matching the \textcolor{red}{classification} pattern  \textcolor{red}{as shown in} Figure~\ref{fig:Decimal_cycle}a(I).
\rev{The extracted sub-GRNN for the Fibonacci classification task is shown in Figure~\ref{fig:Decimal_cycle}a(II), tracing from input-responsive genes through the hidden layer to the output gene.} 


The gene-wise perturbation analysis (Figure~\ref{fig:Decimal_cycle}a(III)) reveals that classification stability is dominated by the most critical gene \textit{b3067}, whereas the majority of hidden-layer genes display low criticality scores (\(<5\)). \textcolor{red}{Therefore, we will focus on gene \textit{b3067} in our analysis.} 
\adrian{Figure~\ref{fig:Decimal_cycle}a(IV) illustrates the effect of perturbing gene \textit{b3067} on the Fibonacci classification sub-GRNN. }
When perturbing the gene \textit{b3067} in input conditions of \(i=1,2\), the output gene expression remains above the threshold across \(\alpha=1\!-\!5\), preserving correct classification.
As \textcolor{red}{the} perturbation increases from \(\alpha = 1\) to \(\alpha = 5\), true positives \(i = 3, 5\) shift below threshold as false negatives, while non-Fibonacci inputs \(i = 4, 6, 7\) cross above \textcolor{red}{the} threshold as false positives.
\adrian{This behavior arises because, under unperturbed conditions, the expression levels of the output gene for inputs \(i = 3, 5\) and \(i = 4, 6, 7\) lie near the decision threshold, making them sensitive to perturbation of \textit{b3067}.}

The collective perturbation analysis in Figure~\ref{fig:Decimal_cycle}a(V) shows that perturbing only the most critical gene (\(k=1\)) raises the Hamming distance to 4 across input codes for perturbation level \(\alpha = 1,2,3\).
For most values of \(k\), the Hamming distance remains in the 2\text{–}3 range and reaches the maximum of 4 only when perturbing \(k = 6,8,9\) number of genes.
\adrian{This behavior arises because the most critical gene $k=1$ occupy key regulatory paths leading to the output gene \textit{b2927}, \textcolor{red}{resulting in} strong propagation of perturbation and causing misclassification. The remaining genes have low criticality (below 5), so their added perturbation has limited impact, leading to saturation ($k \geq 2$) in \textcolor{red}{the} Hamming distance.}
The Lyapunov-based stability under perturbation of gene \textit{b3067} is presented in Figure~\ref{fig:Decimal_cycle}a(VI).
When perturbing the gene \textit{b3067} in input conditions \(i=1,2\), the sub-GRNN demonstrates strong robustness, remaining stable up to higher perturbation thresholds at \(\alpha=7.58\) and \(\alpha=6.22\), respectively.  
However, for input conditions \(i=3,4,5,6,7\), the sub-GRNN becomes unstable at lower perturbation thresholds (\(\alpha=0.63\!-\!1.1\)).
The unperturbed expression levels for input codes \(i = 4,5,6,7\) lie close to the classification threshold, making them highly sensitive to perturbation (see Supplementary Figure~S5a for the Lyapunov stability analysis of the least critical gene and additional results from gene-wise perturbation, which show Hamming distances of 3–4 for many perturbed genes even at low perturbation levels).
The perturbations propagate through the sub-GRNN and elevate the expression above the threshold \(\theta\) in input conditions \(i=1,2\) for lower perturbation levels \(\alpha < 5\).  
This effect decreases the Lyapunov function \(\mathcal{V}(\alpha, \sigma)\) (\ref{eq:classify_lyapunov}).
The minimum perturbation threshold \(\alpha=0.63\), \(\sigma^2=0.17\) across input codes marks the upper bound of tolerable perturbation for stable and reliable computing.

\subsection*{Is the repeating cycle length of \(1/i\) equal to 1?}
\begin{figure*}[t!] 
\centering
\includegraphics[width=\linewidth]{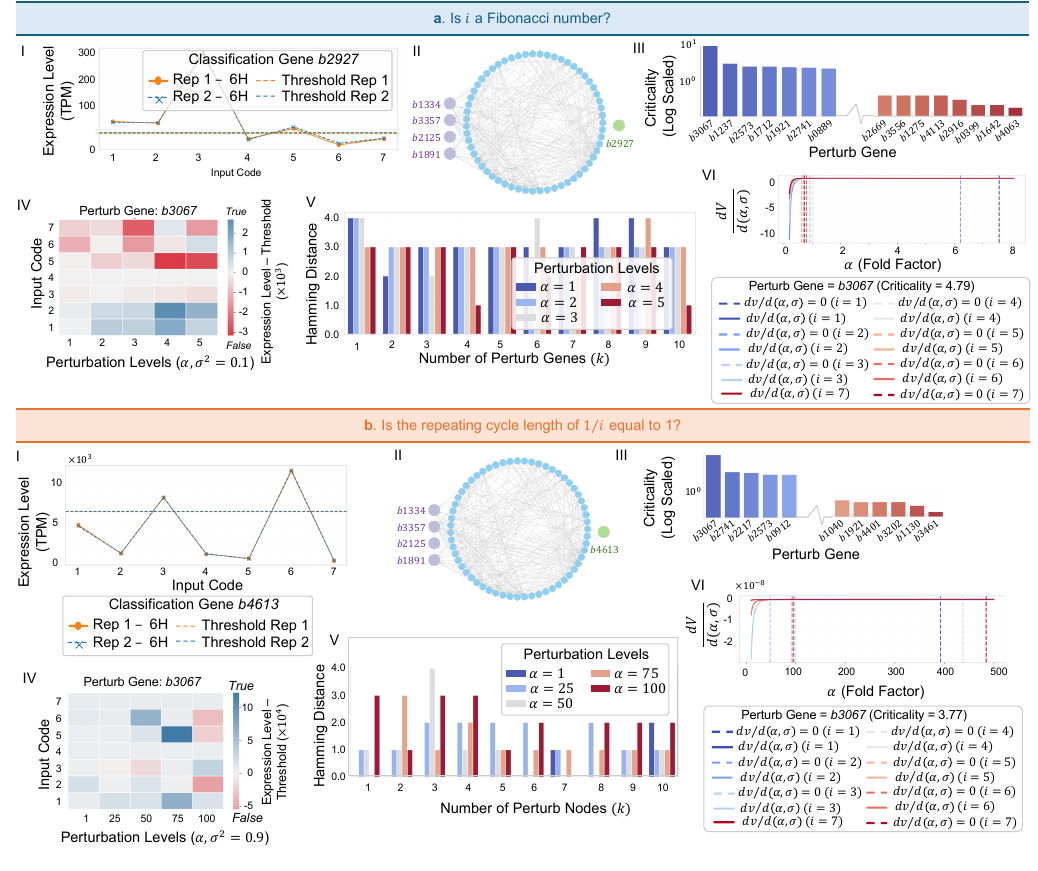} 
\vspace{-2.9em}
\caption{
\textbf{GRNN-based Fibonacci number classification (a) and decimal expansion of $1/i$ have a repeating cycle length of 1 (b).} 
\textbf{a,} Classification of whether input code \(i\) is a Fibonacci number using gene \textit{b2927}. 
\textbf{b,} Classification of whether the decimal expansion of \(1/i\) has a repeating cycle length of 1, using gene \textit{b4613}. 
\textbf{I,} Output expression dynamics across input codes with decision threshold for each replicate. 
\textbf{II,} Sub-GRNN structure linking input genes (purple), hidden-layer regulators genes (blue), and the classifier output genes (green). 
\textbf{III,} The criticality scores under gene-wise perturbation. 
\textbf{IV,} Input-specific perturbation effects on classification error, showing expression deviation from threshold for gene \textit{b3067}. 
\textbf{V,} Collective perturbation impact on classification accuracy measured by Hamming distance across perturbation levels \(\alpha=1, 25, 50, 75, 100\) with fixed noise variance \(\sigma^2=0.9\).  
\textbf{VI,} Lyapunov-based stability analysis for the most critical gene.
}
\vspace{-1em}
\label{fig:Decimal_cycle}
\end{figure*}

\adrian{Among the input codes \(i = 1, 2, \dots, 7\), only \(i = 3\) and \(i = 6\) yield decimal expansions with cycle length 1 in the reciprocal \(1/i\), i.e., \(1/3 = 0.\overline{3}\) and \(1/6 = 0.1\overline{6}\).}  
\adrian{Using the search algorithm for classification tasks (Supplementary Algorithms~2 and 3), we identified gene \textit{b4613} 
as the output gene that matches this binary pattern shown in Figure~\ref{fig:Decimal_cycle}b(I) with the sub-GRNN illustrated in Figure~\ref{fig:Decimal_cycle}b(II). The output pattern for this task exhibited this expression behavior at \(t = 6\) hours.} 

 The gene-wise perturbation analysis (Figure~\ref{fig:Decimal_cycle}b(III)) reveals a criticality distribution with \textit{b3067} showing the highest criticality, followed by a sharp drop to \textit{b2741} and \textit{b2217}, after which the remaining genes decrease gradually, with most falling below a criticality score of 1.
According to the gene-wise perturbation analysis, the decimal cycle length classification sub-GRNN is highly robust to perturbations, requiring very large perturbation amplitudes (\(\alpha \geq 50\)) to observe misclassifications (Figure ~\ref{fig:Decimal_cycle}b(IV)). This perturbation threshold is considerably higher than other sub-GRNN examined in this study.
The perturbation of the most critical gene \textit{b3067} with input codes \(i = 2, 3,5, 6\) preserves correct classification even under high perturbation levels, with misclassifications appearing only at \(\alpha \geq 50\), demonstrating strong robustness.
This behavior is explained by the fact that, for input codes \(i = 2, 3,5, 6\), perturbations to the critical gene \textit{b3067} do not strongly propagate through the sub-GRNN even at higher perturbation levels.
The perturbation of the gene \textit{b3067} in the input conditions corresponding to \(i = 1, 4, 7\) keeps the output gene \textit{b4613} expression consistently below the threshold, ensuring correct classification even under higher perturbation levels.
Figure~\ref{fig:Decimal_cycle}b(V) quantifies the effect of collective perturbations, showing that even when multiple genes are perturbed (\(k = 1\)–10) under very large perturbation amplitudes (\(\alpha \geq 25\)), the Hamming distance remains bounded within 1–4 misclassifications. This demonstrates the strong robustness of the sub-GRNN, as performance remains stable even under extensive collective perturbations.
Figure~\ref{fig:Decimal_cycle}b(VI) depicts Lyapunov stability under perturbation of \textit{b3067}, where input codes \(i=2,3,5,6\) tolerate very high perturbations with instability thresholds at \(\alpha=95.53, 49.59, 98.48,\) and \(92.58\), respectively. In contrast, input codes \(i=1,4,7\) withstand even larger perturbations, with thresholds only at \(\alpha=392.08, 437.25,\) and \(484.39\), demonstrating the strongest robustness among all cases.
This robustness arises because perturbing the most critical gene does not markedly alter the typical expression of output-gene \textit{b4613} in all input conditions, leading to slow growth of \(\mathcal{V}(\alpha, \sigma)\) (\ref{eq:classify_lyapunov}) and demonstrating high tolerance to perturbations.
Therefore, the minimum stability threshold across all inputs is set at \(\alpha = 49.91\), \(\sigma^2 = 0.75\), defining the most conservative upper bound on tolerable perturbation for stable classification.
Supplementary Figure~S5b confirms this sub-GRNN’s robustness by showing that it tolerates very large perturbations (\(\alpha>250\)) while maintaining stable and reliable computing when the least critical gene is perturbed.

\section*{DISCUSSION}

This study presents a bacterial GRNN framework designed to function as a biologically grounded library of mathematical solvers. Motivated by the need for stable and reliable computing in biochemical systems, our approach uses gene expression dynamics to perform mathematical calculation and classification  under different experimental conditions. 
Unlike traditional biocomputing approaches based on engineered logic gates or synthetic circuits, we exploit sub-GRNNs that resemble artificial neural network structures in AI. These sub-GRNNs generate gene expression patterns that match the desired outputs for mathematical tasks such as calculation and classification.  
This work bridges systems biology and natural computing by enabling task-specific sub-GRNN discovery and stability–reliability assessment.

To demonstrate the bacterial GRNN’s versatility as a biocomputing library,
we implement seven mathematical calculation tasks including $i^{th}$ Fibonacci number, Collatz step counting, multiplication, and four classification tasks including detecting prime numbers, Fibonacci numbers, lucky numbers, and repeating cycle length in the decimal expansion of \( 1/i \).
\adrian{Output genes are selected by task-specific expression behavior with fold changes for calculation tasks and binary pattern for classification.}
\adrian{The Collatz step counting task uses a unique step-count encoding and produces more distinct sub-GRNNs than other tasks, reflecting its greater encoding complexity and diversity of expression patterns.}
Expression profiles in $i^{th}$ Fibonacci number and multiplication sub-GRNNs follow exponential and linear trends, respectively, aligning with expected outputs.
Classification tasks exhibit sharper expression transitions, reflecting binary patterns consistent with threshold-based behavior.
Together, the diverse expression patterns across input conditions highlight how the GRNN enables both continuous and discrete computing through problem specific gene regulation patterns.

To assess the stability and reliability of computing and classification tasks of sub-GRNNs, we apply gene-wise perturbation, collective perturbation, and Lyapunov-based stability analyses.  
In mathematical calculation tasks, gene-wise perturbation analysis reveals that a few critical genes, such as \textit{b3067}, have strong topological influence, where small perturbations cause significant output deviations, especially under stress-responsive input conditions. In classification tasks, a few high centrality, upstream genes, such as \textit{b3067}, dominate classifier reliability by disrupting the output-gene whose unperturbed expression levels lie close to the decision threshold of target input codes, increasing  false positives. For non-target input codes, increasing perturbation alters the output gene’s expression above the threshold, altering its typical behavior and causing late false negatives.

Collective perturbation analysis reveals that the sub-GRNN is sensitive to perturbations of a small set of high-criticality genes. In mathematical calculation tasks, coefficient of determination (\(R^2\)) drops rapidly after perturbing 3–5 such genes, while classification tasks show early Hamming distance saturation.
This highlights a modular structure where a core gene set dominates output dynamics, and further perturbations have diminishing effects.
 Lyapunov stability analysis reveals that maximum pertubation level a sub-GRNN can tolerate  decreases with task complexity. Specifically, multiplication and $i^{th}$ Fibonacci tasks show early instability for higher input codes, while classification problems remain stable across a wider range of perturbation levels. Stability also varies across input conditions, with stress-responsive inputs exhibiting earlier instability under the perturbation of critical genes such as \textit{b3067}. This input-specific sensitivity reflects differences in gene expression correlation matrix, which governs perturbation propagation through the sub-GRNN. The minimum stability threshold across input codes defines a conservative upper bound for stable and reliable computing. Among all problem-specific sub-GRNNs, the decimal cycle length sub-GRNN emerges as the most robust to perturbations.


\subsection*{Limitations of the study}
\adrian{A key limitation of this study is that the gene expression patterns matched to computing tasks are specific to the cultured bacterial cells and environmental conditions used in this study, and may not directly generalize across different species or growth contexts.}
To improve model precision, experimental validation of gene-level perturbations, such as CRISPR \cite{Yi2021} knockdowns is needed to confirm the predicted maximum perturbation level for reliable and stable computing.
 Another key direction is integrating control theory \cite{10945666} to systematically determine the optimal chemical input concentration needed to restore  the sub-GRNN output following adversarial or environmental perturbations. Such an approach could turn sub-GRNNs into self-correcting biological circuits that maintain reliable function even under challenging conditions \cite{LEVIN2019519}. The input encoding strategy used in the Collatz solver could be generalized to other complex, nonlinear problems, discovering larger, functionally diverse sub-GRNNs and expanding the biocomputing library. \adrian{This work lays the foundation for stable and reliable biological processors that respond to chemical cues by modulating gene expression, enabling applications in \amr{biosensing \cite{7384676}}, \amr{smart therapeutics \cite{9345800, 9052661}}, and \textit{in vivo} molecular diagnostics \cite{ BOADA2020101305, 9405303}.}









\section*{RESOURCE AVAILABILITY}


\subsection*{Lead contact}


Requests for further information and resources should be directed to and will be fulfilled by the lead contact, Adrian Ratwatte (aratwatte2@huskers.unl.edu).

\section*{ACKNOWLEDGMENTS}


This work was funded by National Science Foundation (NSF) via grants CBET-2316960 and CCF- 2544845. The authors thank all members of the lab for their support.

\section*{AUTHOR CONTRIBUTIONS}


Conceptualization, A.R., S.S., T.C., X.L. and S.B.; methodology, A.R., S.S, T.C., X.L. and S.B.; investigation, A.R., S.S, T.C., X.L. and S.B.; writing original draft, A.R., S.S, T.C., X.L. and S.B.; writing review \& editing, A.R., S.S, T.C., X.L. and S.B.; funding acquisition, S.B.; resources, X.L. and S.B.; supervision, X.L. and S.B.


\section*{DECLARATION OF INTERESTS}


The authors declare no competing interests.

\bibliography{references}

\bigskip


\newpage



\section*{STAR METHODS}

\subsection*{Method details}

This section outlines the experimental and computational workflow used in this study. 
We begin with discussing the problem-specific sub-GRNN search algorithm, followed by perturbation and Lyapunov stability analyses. The Supplementary  details transcriptomic data collection, GRN inference from publicly available datasets and the correlation analysis of gene expression. 

For the correlation analysis, we used five publicly available time-series gene expression datasets.
For each dataset, we calculated the Pearson correlation between every regulator–target gene pair in the GRN, yielding five independent correlation values for each edge. To determine  stable edges across these input conditions, we defined a consistency score (Equation~S1) based on the proportion of correlations matching the overall mean sign, scaled by the inverse of their standard deviation across datasets.
The consistency score gives higher scores to edges that stay consistent across conditions and lower scores to edges that change significantly.  The denominator reduces the score for unstable edges, and the final value is kept between 0 and 1, where values closer to 1 imply the edge is stable. 
Across chemical conditions, the correlation analysis shows that about \(30\%\) of edges are stable, whereas approximately  \(70\%\) vary with the input condition, indicating dynamic regulatory responses. For the deitailed explanaiton  see Supplementary Note 2.


\subsubsection*{Problem-specific Sub-GRNN Search Algorithm}
This section outlines the algorithms used to extract sub-GRNNs for mathematical calculation tasks and classification.
For mathematical calculation tasks, we consider the fold-change of each gene, defined as the ratio between its expression level under a given condition and the baseline condition. 
To identify genes exhibiting  fold-change patterns aligned with predefined expected pattern, we use a two-step approach (Algorithm 1 in  Supplementary). First, each gene is evaluated across conditions and time points, verifying that the observed fold-changes in both replicates remain within a tolerance \(\epsilon\) of the expected values. Finally, the best matching  gene and the corresponding time point $(p^*, t^*)$ are chosen as the output by minimizing the total deviation $\delta$, ensuring the closest match to the desired calculation behavior. 

To identify output genes suitable for classification tasks, we apply a threshold-based selection procedure that determines  genes and the corresponding time points for consistent agreement with a predefined binary pattern across replicates (Step 1; see Algorithm 2 in Supplementary). Among the selected output-genes, we then assess expression separation by computing the distances of expression values from a gene-specific threshold, identifying the closest values above and below it in each input code to compute midpoint thresholds that define the decision boundary (Step 2: see Algorithm 3 in Supplementary). A total score is then calculated as the sum of these distances across both target and non-target input codes. The gene and time point with the lowest total score, showing the clearest separation between target and non-target input codes is selected as the output-gene.

The Collatz step count problem uses a distinct binary encoding approach, searching output genes that exhibit consistent  binary 
expression patterns across conditions. 
To identify such genes, we evaluate each gene and time point for threshold-based agreement with a predefined binary pattern in both replicates (Algorithm~4, Supplementary).  Thresholds are defined by the midpoint of the largest adjacent expression gap, and only genes that reproduce the expected binary activation across replicates are retained.  For each gene and time point, we measure the separation between target from non-target inputs. We then select the gene and time point with the largest separation and use its midpoint as the decision threshold. 

\subsubsection*{Perturbation Analysis}
To evaluate the stability and reliability of computing under gene perturbations, we develop a gene-wise perturbation analysis (Supplementary Algorithms~5, 6). 
This analysis simulates perturbation propagation  of a single gene \(p\) through its downstream neighbors using a correlation-weighted adjacency matrix \(W\), thereby impacting the output gene and the sub-GRNN computing.  Additive perturbations \(\mathbf{u}_t\), sampled from a zero-mean Gaussian distribution and scaled by fold factor \(\alpha\) and noise variance \(\sigma^2\), are propagated through \(W[p,:]\) to yield perturbed gene expression matrices \(\tilde{\mathcal{X}}^{(\alpha)}_{c,r}\). This formulation identify the critical regulatory genes whose perturbations induce expression changes in output-gene.

To quantify the stability and reliability of GRNN-based mathematical calculation under gene-wise perturbations, we evaluate the deviation between expected $\Delta^{(0)}_{\text{expected}}$ and perturbed $\Delta^{(\alpha, \sigma^2)}_{\text{actual}}$ output fold-changes following perturbation of gene \(p\).
 For each perturbation level \(\alpha\) and $\sigma^2$, the actual fold change and corresponding error $\varepsilon$ are computed as:
\begin{align}
\Delta^{(\alpha, \sigma^2)}_{\text{actual}} &= \frac{x^{(\alpha, \sigma^2)}_{c}}{x^{(\alpha, \sigma^2)}_{i=base}} \\
\varepsilon^{(\alpha, \sigma^2)} &= \left| \Delta^{(0)}_{\text{expected}} - \Delta^{(\alpha, \sigma^2)}_{\text{actual}} \right|,
\end{align}
where \(\Delta^{(\alpha, \sigma^2)}_{\text{actual}}\) denotes the calculated fold change in gene expression under perturbation level \(\alpha\) and $\sigma^2$, computed as the ratio between the perturbed expression \(x^{(\alpha, \sigma^2)}_c\) at input code \(i\) and the expression in baseline input condition \(x^{(\alpha, \sigma^2)}_{i=base}\). 
To quantify the overall deviation across perturbation levels, we compute the Error Sum of Squares (ESS) for perturbation gene \(p\), where \(\varepsilon^{(\alpha, \sigma^2)}\) denotes the error:
\begin{align}
\text{ESS}_p = \sum_{\alpha, \sigma^2} \left( \varepsilon^{(\alpha, \sigma^2)} \right)^2. \label{eq:ESS}
\end{align}
To assess the variability of unperturbed expression levels $x$, we compute the expression variance of the output gene \(g_{\text{out}}\) across input codes \(i\) as follows:
\begin{align}
\text{Var}_{g_{\text{out}}}^{(0)} = \sum_{i} \left( x^{(0)}_{i, g_{\text{out}}} - \bar{x}^{(0)}_{g_{\text{out}}} \right)^2, \label{eq:expression_variance}
\end{align}
where $x^{(0)}_{i, g_{\text{out}}}$ is the unperturbed expression level of output gene $g_{\text{out}}$ in input code $i$, and $\bar{x}^{(0)}_{g_{\text{out}}}$ is the mean expression level of $g_{\text{out}}$ across all input codes. 
By applying \ref{eq:ESS} and \ref{eq:expression_variance}, we compute the coefficient of determination \(R^2\) as follows:
\begin{align}
R^2_{p,\alpha,\sigma^2} = 1 - \frac{\text{ESS}_{p,\alpha,\sigma^2}}{\text{Var}_{g_{\text{out}}}^{(0)}}.\label{eq:r_squared}
\end{align}
The mean \(R^2\) across perturbation levels is given by:
\begin{align}
\overline{R^2}_p = \frac{1}{|\boldsymbol{\alpha, \sigma^2}|} \sum_{\alpha, \sigma^2} R^2_{p,\alpha, \sigma^2}, \label{eq:mean_r_squared}
\end{align}
where \(|\boldsymbol{\alpha, \sigma^2}|\) is the pertubation levels applied, and \(R^2_{p,\alpha, \sigma^2}\) denotes the coefficient of determination for gene \(p\) under perturbation level \(\alpha\) and $\sigma^2$.

To quantify a gene’s impact on computing stability and reliability, we define its criticality \(C_p\) as the ratio of outward degree centrality to its average coefficient of determination \(R^2\), capturing how strongly perturbations can spread through the sub-GRNN and influence output expression:
\begin{align}
C_p = \frac{\text{Outward Centrality}_p}{\overline{R^2}_p}. \label{eq:criticality}
\end{align}

For classification problems we define criticality using the \textit{Hamming distance}, which measures the number of mismatches between the unperturbed binary output \(y_i\) which is assigned based on unperturbed expression levels and the perturbed output \(\hat{y}^{(\alpha)}_i\) across all input codes \(i\):
\begin{align}
\text{HD}_{p,\alpha} &= \sum_{i=1}^{N} \mathbb{1}\left[ y_i \neq \hat{y}^{(\alpha)}_i \right], \label{eq:hamming}
\end{align}
where \(\mathbb{1}[\cdot]\) is the indicator function, and \(N\) is the number of input codes.
The classification-based criticality \(C^{\text{class}}_p\) is then defined as the product of the gene’s outward degree centrality and the total Hamming distance across all perturbation levels \(\alpha\):
\begin{align}
C^{\text{class}}_p &= \text{Centrality}_p \cdot \sum_{\alpha} \text{HD}_{p,\alpha}. \label{eq:criticality_class}
\end{align}

To evaluate the global impact of biologically plausible perturbations, we extend our framework to collective perturbations (see Supplementary Algorithm~6). Unlike the gene-wise perturbation, this approach models how simultaneous perturbation of a set of genes \(\mathcal{P} \subseteq V\), as occurs with chemical stimuli, propagates through the sub-GRNN. Perturbations \(\mathbf{u}_t\), sampled independently for each \(p \in \mathcal{P}\), are scaled by fold factor \(\alpha\) and noise variance \(\sigma^2\), then propagated via \(W[p,:]\), with their cumulative influence yielding the perturbed expression matrix \(\tilde{\mathcal{X}}^{(\alpha)}_{c,r}\).


\subsubsection*{Stability Analysis using Lyapunov Stability Theorem}

In this section, we describe the Lyapunov stability theorem used to determine the maximum per-gene perturbation a sub-GRNN can tolerate while maintaining stable and reliable computing.

We define a Lyapunov function based on gene-wise perturbation and establish its stability conditions.
Let \( G = (V, E) \) be the GRN, where \( V \) is the set of \( n \) genes and \( E \) is the set of directed regulatory edges. Let \( x^*(t) \in \mathbb{R}^n \) 
denote the unperturbed gene expression levels at time \( t \), and \( x(t; \alpha, \sigma) \in \mathbb{R}^n \) the perturbed gene expression levels, where \( \alpha > 0 \) represents the fold-change amplitude relative to the unperturbed maximum and minimum gene expression levels, and \( \sigma \in \mathbb{R}^n \) captures the variance in the perturbed expression levels. 
The correlation matrix \( \mathbf{W} \) was constructed from publicly available transcriptomic datasets by computing pairwise correlations of gene expression levels, yielding coefficients that quantify gene–gene interaction influences within the GRNN.
In the  matrix \( \mathbf{W} \in \mathbb{R}^{n \times n} \), each row \( p \) corresponds to gene \( p \) as the source-gene, each column corresponds to a target gene, and \( W_{pp} = 1 \) for all \( p \). The correlation matrix is described in detail in Supplementary Note~4.1 and Fig.~S2.  
The perturbed gene expression levels of the sub-GRNN is given by
\begin{align}
    \mathbf{x}(t; \alpha, \sigma) = \mathbf{x}^*(t) + \mathbf{u}(t; \alpha, \sigma) \mathbf{W}, \label{eq:state_eqn}
\end{align}
where \( \mathbf{x}^*(t) \in \mathbb{R}^n \) is the unperturbed expression vector, \( \mathbf{W} \in \mathbb{R}^{n \times n} \) is the correlation matrix, and \( \mathbf{u}(t; \alpha, \sigma) \in \mathbb{R}^n \) is the perturbation vector defined as
\begin{align}
    \mathbf{u}(t; \alpha, \sigma) = \alpha \, \mathrm{diag}(\mathbf{x}^{\max} - \mathbf{x}^{\min}) \, (\boldsymbol{\sigma^2} \odot \boldsymbol{\eta}), \label{eq:noise_eqn}
\end{align}
where \( \mathbf{x}^{\max} \) and \( \mathbf{x}^{\min} \) are the vectors of maximum and minimum unperturbed expression levels, \( \boldsymbol{\sigma} \) is the standard deviation vector, \( \boldsymbol{\eta} \) is a random vector with entries drawn from a Gaussian distribution, and \( \odot \) denotes element-wise multiplication.
Let $ \delta \textbf{x} = x- x^* $ then $ \delta \textbf{x} = \mathbf{u}(t; \alpha, \sigma) \mathbf{W}$ from \ref{eq:state_eqn}. Suppose each gene \( p \in V \) experiences a perturbation \( \delta x_p \) modeled as
\begin{align}
    \delta \textbf{x}_p = \alpha (x_p^{\max}(t) - x_p^{\min}(t)) \sigma_p^2 \eta_p \left( W_{p1}, W_{p2}, \dots, W_{pn} \right). \label{eq: perturbation_added}
\end{align}
where \( \alpha > 0 \) is a fold factor, \( (x_p^{\max} - x_p^{\min}) \) denotes the fold-change range of gene \( p \), \( \sigma_p > 0 \) represents the noise variance of gene \( p \), and \( \eta_p \sim \mathcal{N}(0,1) \) is a standard Gaussian variable modeling zero-mean random fluctuations.

We define the Lyapunov  function for the mathematical calculation problems as:
\begin{align}
\mathcal{V}(i, t; \alpha, \sigma) 
&= \frac{\mathcal{C}(p)^2}{1 + \mathcal{C}(p)^2} 
 \sum_{q \in \mathcal{N}^+(p)} \left\| 
 \frac{w_q(\alpha, \sigma) \delta x_q(i, t)}{\alpha_q^{3/2} \sigma_q \left(1 + \delta x_q(i, t)\right)} 
\right\|^2,
\label{eq:Lyapunov_fn}
\end{align}
\begin{align}
    w_q(\alpha, \sigma) &= 1 + \cdot \max\left(0, \frac{ \delta x_q(i, t; \alpha, \sigma) - \delta x_q(i, t; \alpha_0, \sigma_0) }{ \delta x_q(i, t; \alpha_0, \sigma_0) + \zeta } \right), \label{eq:weight_lyapunov}
\end{align}
where \( \mathcal{C}(p) \) denotes the criticality score of gene \( p \), reflecting its significance from gene-wise perturbation analysis and outward degree centrality, 
\( \mathcal{N}^+(p) \) is the set containing gene \( p \) and its neighboring genes, 
\( \alpha_p, \alpha_q > 0 \) are fold factors for genes \( p \) and \( q \), respectively,  
\( \sigma_p, \sigma_q > 0 \) represent noise variances at genes \( p \) and \( q \),  
\( \delta x_q(i, t) \) is the expression deviation of the output gene at node \( q \) under the perturbation for input code \( i \) and time \( t \), 
\( w_q(\alpha, \sigma) \) is a weight factor quantifying the relative increase in \( \delta x_q \) compared to the minimal perturbation levels \( (\alpha_0, \sigma_0) \) within the perturbation range, with \(\zeta > 0\) preventing division by zero.

For classification problems, the Lyapunov function is formulated as follows.
Let $x^*$ denote the unperturbed gene expression level, and $x(\alpha)$ the perturbed expression level under perturbation level $\alpha$. Let the decision threshold be $\theta$, and define:
\begin{align}
    y &= \mathbf{1}[x_\text{start} > \theta], \quad \text{(true label)} \\
    \hat{y}(\alpha) &= \mathbf{1}[x(\alpha) > \theta], \quad \text{(predicted label)} \\
    r(\alpha) &= \mathbf{1}[\hat{y}(\alpha) \ne y], \quad \text{(misclassification indicator)} \\
    \text{dist}(\alpha) &= |x(\alpha) - \theta|, \quad \text{(distance to threshold)} \\
    \text{sign}_{\text{err}}(\alpha) &= \text{sign}\left( (x(\alpha) - \theta)(x_\text{start} - \theta) \right),
\end{align}

where a misclassification occurs when the perturbed expression level and its expected binary output \(\hat{y}(\alpha)\) differs from the actual binary output \(y\), and the sign term indicates whether the perturbation crosses the decision boundary.
To identify input codes that are more likely to flip classification under small changes, we define a directional penalty term:
\begin{align}
    \beta(\alpha) = \left(1 - \text{sign}_{\text{err}}(\alpha)\right) \cdot \frac{1}{1 + \text{dist}(\alpha)}.
\end{align}
The term $\beta(\alpha)$  becomes larger when the perturbed expression crosses the threshold from the opposite side and is close to it, highlighting  misclassified  input codes which are closer to the decision boundary.

Then, the Lyapunov  function for the classification problems is given by:

\begin{align}
    \mathcal{V}(\alpha, \sigma) 
    &= \frac{\sigma^2 \mathcal{C}(p)^2}{\alpha(1 + \mathcal{C}(p)^2)} 
    \sum_{q \in \mathcal{N}^+(p)} 
    \left\| 
    \frac{w_q(\alpha, \sigma)}{\alpha_q^{1/2} \sigma_q}
    \left(   \beta_q +  r_q \right)
    \right\|^2, \label{eq:classify_lyapunov}
\end{align}
where
\begin{align}
w_q(\alpha, \sigma) 
= 1 +  \cdot \max\left( 0, \frac{\text{dist}_q(\alpha) - \text{dist}_q^{\text{base}}}{\text{dist}_q^{\text{base}} + \zeta} \cdot r_q(\alpha) \right).
\end{align}
Here, the indicator \( r_q \) denotes misclassification, \( \beta_q \) is the directional deviation toward incorrect classification, \( \text{dist}_q(\alpha) \) is the distance between the perturbed expression, \(\zeta>0\) is a small constant to avoid division by zero and the decision threshold \( \theta \), and \( \text{dist}_q^{\text{base}} \) is the  distance between the perturbed expression under minimum perturbation within the given perturbation range  and the decision threshold.

The Lyapunov functions in this study are unique for the GRNN-based computing because it captures the average neighborhood error propagation per perturbation level between a gene \( p \) and its neighbors \( q \), reflecting the distributed and interconnected structure of the sub-GRNN. By summing the perturbation-induced deviations of both \( p \) and \( q \), the function measures the cumulative deviation from the steady state when the gene and its local neighborhood are perturbed. A lower \( \mathcal{V}_p \) value implies that \( p \) and its neighboring genes are robust to perturbations, whereas a higher value indicates that \( p \) and its neighboring genes are highly sensitive to perturbation.

To verify the Lyapunov conditions, note that \( \mathcal{V}(t; \alpha, \sigma) \) is a finite sum of squared norms, and thus \( \mathcal{V}(t; \alpha, \sigma) \geq 0 \) for all \( t \), establishing positive semi-definiteness \cite{zhou2015sufficient}.
    Furthermore, \( \mathcal{V}(t; \alpha, \sigma) = 0 \) if and only if \( \delta x_p = 0 \) and \( \delta x_q = 0 \) for all \( (p,q) \in E \).
By substituting (\ref{eq: perturbation_added}) into (\ref{eq:Lyapunov_fn}), we obtain:
\begin{align}
    &\mathcal{V}(t; \alpha, \sigma) 
    = \frac{\sigma^2 \mathcal{C}(p)^2}{ \alpha (1+ \mathcal{C}(p)^2)} 
    \sum_{q \in \mathcal{N}^+(p)} 
     \left\|  
  \frac{w_q(\alpha, \sigma) \, \alpha_q (x_q^{\max}(t) - x_q^{\min}(t)) \sigma_q^2 \eta_q \left( W_{q1}, W_{q2}, \dots, W_{qn} \right)}{\alpha_q^{3/2} \sigma_q \left(1 + \alpha_q (x_q^{\max}(t) - x_q^{\min}(t)) \sigma_q^2 \eta_q \left( W_{q1}, \dots, W_{qn} \right) \right)} 
\right\|^2,
\end{align}
where, for simplicity, we assume \( \alpha = \alpha_p = \alpha_q \) and \( \sigma = \sigma_p = \sigma_q \). 
Here, \( \mathcal{C}(p) \) is the criticality score of gene \( p \), \( \mathcal{N}^+(p) \) the set of \( p \) and its neighbors, $n$ is the number of neighboring genes of p, \( w_q(\alpha, \sigma) \) is the weight factor, \( x_q^{\max}(t) \) and \( x_q^{\min}(t) \) are the maximum and minimum unperturbed expression levels of gene \( q \), \( \sigma_q \) is the noise variance, \( \eta_q \sim \mathcal{N}(0,1) \) a Gaussian random variable, \( \alpha_q > 0 \) the fold-change factor, and \( W_{qj} \) is the correlation coefficient between genes \( q \) and \( j \).  
For the mathematical calculation problems we define:
\begin{align}
    \Delta_{pq} 
    &:= \frac{w_q(\alpha, \sigma) \,  (x_q^{\max}(t) - x_q^{\min}(t))  \eta_q \left( W_{q1}, W_{q2}, \dots, W_{qn} \right)}{  \left(1 +  (x_q^{\max}(t) - x_q^{\min}(t))  \eta_q \left( W_{q1}, \dots, W_{qn} \right) \right)} ,
\end{align}
and for the classification problems Lyapunov function from (\ref{eq:classify_lyapunov}) becomes:
\begin{align}
    \Delta_{pq} 
    &:= \frac{w_q(\alpha, \sigma)}{\alpha_q^{1/2} \sigma_q}
    \left(   \beta_q +  r_q \right),
\end{align}
where \( w_q(\alpha, \sigma) \) is the weight factor, \( x_q^{\max}(t) \) and \( x_q^{\min}(t) \) are the maximum and minimum unperturbed expression levels of gene \( q \), \( \eta_q \sim \mathcal{N}(0,1) \) is a Gaussian random variable, \( W_{qj} \) are correlation coefficients from the matrix \( \mathbf{W} \), \( \alpha_q > 0 \) is the fold factor, \( \sigma_q \) the noise variance, \( r_q \) the misclassification indicator, and \( \beta_q \) the directional deviation toward incorrect classification.

Therefore, the Lyapunov function for both mathematical calculation and classification tasks becomes:
\begin{align}
    \mathcal{V}(t; \alpha, \sigma) = 
\left( \epsilon + \frac{\sigma^2 \mathcal{C}(p)^2}{ \alpha (1+\mathcal{C}(p)^2) } \right)  \sum_{q \in \mathcal{N}^+(p)} \left\| \Delta_{pq} \right\|^2. \label{eq: Lyapunov_fn_compact}
\end{align}

To compute the evolution of \( \mathcal{V} \) with respect to the perturbation level \( s \), we apply the total derivative:
\begin{align}
    \frac{d\mathcal{V}}{ds}
    &= l \frac{\partial \mathcal{V}}{\partial \sigma} + \frac{k}{\alpha^2} \frac{\partial \mathcal{V}}{\partial \alpha}  \\
    &= \frac{\sigma \mathcal{C}(p)^2}{ \alpha(1+ \mathcal{C}(p)^2)} \left( 2l - \frac{k \sigma}{ \alpha^3} \right) \sum_{q \in \mathcal{N}^+(p)} \left\| \Delta_{pq} \right\|^2. \label{eq:Lyapunov_derivative}
\end{align}

where \( \alpha = \alpha(s) \) and \( \sigma = \sigma(s) \) are parameterized by \( s \) as:
\begin{align}
    \alpha(s) &= \alpha_0 + k s, \\
    \sigma(s) &= \sigma_0 + l s,
\end{align}
with positive constants \( k \) and \( l \), and \(\gamma(s) = (\alpha(s), \sigma(s))\).
We set \(\alpha_0 = 0.1\), \(\sigma_0 = 0.1\), with \(k = 10\) and \(l = 1\) in the simulations throughout this study to define the perturbation trajectory \(\gamma(s)\).


To determine the maximum perturbation level a sub-GRNN can tolerate under perturbation of a gene while maintaining stable and reliable computing, we analyze the Lyapunov derivative in \ref{eq:Lyapunov_derivative}. When \( \tfrac{dV}{ds} < 0 \), the sub-GRNN tends toward stability, where stability corresponds to remaining in a steady state with the output expression  close to the expected pattern, and when \( \tfrac{dV}{ds} \geq 0 \), it tends toward instability. In contrast to empirical error-based metrics, which require arbitrary thresholds to define when error becomes significant, the Lyapunov approach provides a theoretical and continuous assessment of the sub-GRNN stability across all perturbation levels.
 The critical perturbation level \( s_1 \) at which the derivative vanishes, indicating the transition to instability of computing, is given by
\begin{equation}
s_1 = 1.23 \left( A^{1/3} + \frac{1}{k^2 A^{1/3}} \right) - \frac{0.1}{k \left\| \Delta_{q} \right\|}, 
\end{equation}
where
\begin{equation}
A = 
\sqrt{\frac{0.03}{k^4 l^2} - \frac{0.02}{k^5 l^2} + \frac{1}{k^6}}
- \frac{0.12}{k^2 l} + \frac{0.12}{k^3}.
\end{equation}

\end{document}


\maketitle

\tableofcontents

\suppsection{Transcriptomic Data Collection and Pre-processing} \label{sec:data_collect_preprocess}

In this section, we describe the transcriptomic data collection and preprocessing steps used to generate reliable gene expression profiles for input codes \( i = 1, \dots, 7 \) for downstream analysis.

\suppsubsection{Transcriptomic Data Collection}\label{sec:data_collect}
\textit{Escherichia coli} strain K12 cells were grown in different growth media. For Input Code~1 (High Glucose), \textit{E.~coli} was grown in an M9 minimal medium amended with 6~mM glucose. For Input Code~2 (Low Glucose), \textit{E.~coli} was grown in an M9 medium amended with 0.1~mM glucose. For Input Code~3 (High Glucose + Mn), \textit{E.~coli} was grown in an M9 medium amended with 6~mM glucose and 0.5~mM MnCl$_2$. For Input Code~4 (Nitrate), \textit{E.~coli} was grown in a 20X diluted LB broth amended with 5~mM NaNO$_3$. For Input Code~5 (Nitrate + Mn), \textit{E.~coli} was grown in a 20X diluted LB broth amended with 5~mM NaNO$_3$ and 0.5~mM MnCl$_2$. For Input Code~6 (Low Tryptophan + High Lactose), \textit{E.~coli} was grown in an M9 medium amended with 0.1~mM tryptophan and 3~mM lactose. For Input Code~7 (High Tryptophan + Low Lactose), \textit{E.~coli} was grown in an M9 medium amended with 1~mM tryptophan and 0.3~mM lactose. For all \textit{E.~coli} cultures, they were grown at 30\,$^\circ$C on a shaker. Bacterial samples were collected from the culture at 6, 30, and 70~hours after the onset of cultivation. RNA was extracted from the cell culture and sent to the University of Nebraska Medical Center Genomic Core for sequencing. The transcriptomic data were subjected to bioinformatic analyses.

\suppsubsection{Data Pre-processing}\label{sec:pre_process}
This preprocessing workflow closely follows established RNA‑seq pipelines integrating Trimmomatic, FastQC, HISAT2, and featureCounts \cite{pertea2016transcript}. First, we use \texttt{Trimmomatic} (v0.39) with the TruSeq3 adapter to remove sequencing adapters and low-quality bases, ensuring that only reliable reads are used for the study. 
We run \texttt{FastQC} on both raw and trimmed reads to check sequencing quality before and after trimming. The high-quality reads are then aligned to the \textit{Escherichia coli} K-12 reference genome using \texttt{HISAT2} (v2.2.1), so that each read is mapped to its correct position in the genome. The \texttt{featureCounts} assigns the aligned reads to genes based on the reference annotation, giving us gene-level counts for expression analysis.  

Finally, a custom Python script processes the \texttt{featureCounts} output to calculate gene expression in Transcripts Per Million (TPM). For each sample, it takes the gene identifiers, transcript lengths, and raw read counts. It then divides the counts by transcript length to get Reads Per Kilobase (RPK), and normalizes these values by the total RPK across all genes to compute TPM. 

\suppsection{Stable Edges Analysis} \label{sec:stable_edge_analysis}

To determine stable edges in the \textit{E.~coli} GRN, we examined how consistent the regulatory edges and gene expression levels remain across different input conditions using correlation-based analysis \cite{mpindi2016consistency, guo2021exploring}. This allows us to identify reliable subnetworks with stable edges that can function as robust computing units.  



We used five publicly available time-series gene expression datasets, including GSE65244 and four additional GEO series ranging from GSM175175161 to GSM314206 \cite{doi:10.1073/pnas.0510995103}, \cite{edgar2002gene}.
For each dataset, we calculated the Pearson correlation between every regulator–target gene pair in the GRN. This gave us five separate correlation values for each edge, one for each experimental condition. To quantify stable edges across these input conditions, we defined a consistency score based on the proportion of correlations that match the overall mean sign, scaled by the inverse of their standard deviation across datasets.
The resulting metric is given by:
\begin{align}
\text{Consistency Score} 
= \frac{\text{Same Sign Proportion}}{1 + \text{Standard Deviation of Correlation}},
\end{align}
where the \textit{Same Sign Proportion} is the fraction of correlation values that have the same sign as the mean correlation across all input conditions, and the Standard Deviation of Correlation quantifies the variability of those values. This formulation gives higher scores to edges that stay consistent across conditions and lower scores to edges that change significantly.  The denominator reduces the score for unstable edges, and the final value is kept between 0 and 1, where values closer to 1 imply the edge is stable.  
\begin{figure}[t!] 
\centering
\includegraphics[width=0.9\linewidth]{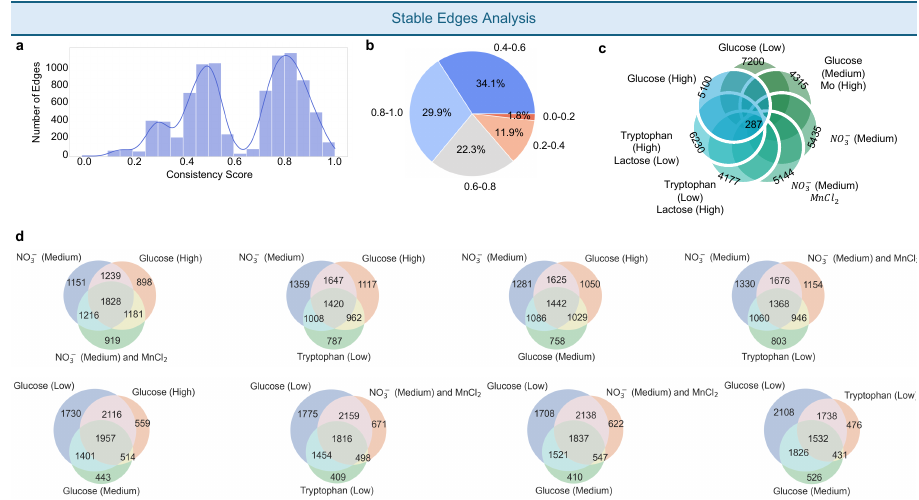} 
\vspace{-1em}
\caption{\textbf{Stable edge analysis across input conditions.} 
\textbf{a,} Distribution of consistency scores for gene--gene regulatory edges, aggregated across five public transcriptomic datasets. 
\textbf{b,} Pie chart showing the percentage of edges within the consistency scores. 
\textbf{c,} Venn diagram illustrating the overlap of the 30\% stable edges across the input conditions used in this study. 
\textbf{d,} Three-way Venn diagrams showing pairwise and triplet overlaps among selected condition subsets from panel c.
}
\vspace{-1.5em}
\label{fig:stable_edges_analysis}
\end{figure}

Figure~\ref{fig:stable_edges_analysis}a shows the distribution of consistency scores for GRN edges across five transcriptomic datasets with different input conditions. The histogram has two peaks: one around moderate values (0.5–0.6) and another, sharper peak at higher values (0.8–0.9), showing that a subset of regulatory edges remains consistently stable across conditions.  
Edges with low consistency scores (\(<0.4\)) show high variability in gene expression across conditions, whereas edges with scores above 0.75 are stable edges and can be used as reliable components for computing.  
Approximately 30\% of the edges fall in the stable range, indicating the presence of a stable regulatory core. These stable edges provide a foundation for exploring sub-GRNNs within a large search space that can be used for  reliable and stable computing \cite{10945666, SOMATHILAKA2023100118, RATWATTE2024100158}.  
The pie chart (Figure~\ref{fig:stable_edges_analysis}b) shows the distribution of edge consistency scores. More than 60\% of edges have scores above 0.6, with 29.9\% in the 0.8–1.0 range and 34.1\% in the 0.4–0.6 range, while only 1.8\% fall below 0.2. These results indicate that most regulatory interactions remain consistent across transcriptomic conditions.

To evaluate how the globally stable edges identified in the top 30\% of the consistency analysis are distributed across the experimental conditions, we examined their overlap across seven transcriptomic inputs (Figure~\ref{fig:stable_edges_analysis}c). The highest number of stable edges was found under low glucose with 7,204 edges, followed by high tryptophan with 6,231 edges and high glucose with 5,146 edges, showing that carbon source availability strongly influences conserved gene–gene correlations. Chemically diverse stress conditions such as \textit{NO$_3^-$ (Medium)} and \textit{MnCl$_2$} retain more than 5{,}000 stable edges.
As more input conditions are considered together, the number of common stable edges decreases. On average, a single condition contains more than 5,300 stable edges, but this drops to about 3,200 when two conditions are intersected and to about 1,900 when three conditions are intersected.  
The central intersection contains 287 gene–gene interactions preserved across all seven conditions, representing the most stable edges under diverse inputs.  
The three-way Venn diagrams (Figure~\ref{fig:stable_edges_analysis}d) show that the largest shared subsets appear when glucose is one of the input conditions, suggesting that glucose regulation contributes disproportionately to stability. When nitrate or tryptophan inputs are included, the overlaps become smaller.

\suppsection{ Problem-specific Sub-GRNN Search Algorithm }\label{sec:search_algo}
In this section, we describe the search algorithms used to identify output genes of sub-GRNNs for both mathematical calculation and classification tasks.  Table~\ref{tab:alg_vars} summarizes the notation used in Algorithms~\ref{alg:calculation}--\ref{alg:collective_perturbation} and provides concise descriptions of each symbol.

\begin{table}[t!]
\centering
\caption{Variables used in Algorithms~\ref{alg:calculation}--\ref{alg:collective_perturbation} and their descriptions.}
\label{tab:alg_vars}
\begin{tabular}{|l|p{13cm}|}
\hline
\textbf{Symbol} & \textbf{Description} \\
\hline
\(x^{(r)}_{(i,p,t)}\) & Expression of gene \(p\) under input code \(i \in \mathcal{I}\) at time \(t\), replicate \(r \). \\
\(\mathcal{I}\) & Set of input codes (experimental conditions). \\
\(V\) & Set of genes in the GRN. \\
\(\mathcal{F}\) & Expected fold-change or classification pattern. \\
\(\epsilon\) & Tolerance for matching fold-change patterns. \\
\(\mathcal{V}_{\mathrm{Match}}\) & Set of ouput-genes that matches the expected pattern. \\
\(d^{(r)}_{(i,p,t)}\) & Observed fold-change \(x^{(r)}_{(i,p,t)}/x^{(r)}_{(i_0,p,t)}\) for the base input code \(i_0\). \\
\((p^{*},t^{*})\) & Best-matching output gene and time point. \\
\(\theta_{p}\) & Mean gene expression over input codes for gene $p$. \\
\(\hat{\theta}^{(r)}_{p}\) & Midpoint thresholds. \\
\(\delta^{(r)}_{\text{above}}, \delta^{(r)}_{\text{below}}\) & Closest distances between gene expression levels and the threshold. \\
\(\theta^{(r)}\) & Threshold from the largest adjacent expression gap. \\
\(x^{+},x^{-}\) & Closest expression values just above and just below \(\theta^{(r)}\). \\
\(\rho(i,j)\) & Correlation weight on edge \((i,j)\). \\
\(D_{\max}\) & Maximum perturbation propagation depth through the sub-GRNN.\\
\(\Pi_{(i,r)}(j)\) & Set of path-wise cumulative correlation products from perturb gene to output-gene \(j\). \\
\(W_{(i,r)}\) & Correlation-weighted matrix.. \\
\(x_{\min}(p),\,x_{\max}(p)\) & Min/max unperturbed expression of gene \(p\) across input codes. \\
\(\mathbf{u}_{t}\) & Sampled perturbation vector at time \(t\). \\
\(\tilde{x}^{(r)}_{(i,p,t)}\) & Perturbed expression matrix. \\
\hline
\end{tabular}
\vspace{-1.4em}
\end{table}

\begin{algorithm}[t!]
\caption{Sub-GRNN Search Algorithm for Mathematical Calculation Problems.}
\label{alg:calculation}

\KwIn{
\begin{itemize}
  \item Gene expression levels \(x^{(r)}_{(i, p, t)}\) of $p^{th}$ gene corresponding to  $i^{th} \in \mathcal{I}$ input code  at time $t$,  for replicate \(r = 1,2\).
  \item Input codes  \(\mathcal{I} = \{1, 2, \ldots, 7\}\).
  \item Gene set \(V\) in the GRN.
  \item Expected fold-change map \(\mathcal{F}: \mathbb{Z}^+ \rightarrow \mathbb{R}^+\).
  \item Tolerance \(\epsilon > 0\).
\end{itemize}
}
\KwOut{
Matched output-genes \(\mathcal{V}_{\mathrm{Match}}\); Matched output-gene and the time point \((p^*, t^*)\)
}

\Begin{
\textbf{Step 1: Find Matching Output-gene and the Time Point Across Replicates} \\
Initialize: \(\mathcal{V}_{\mathrm{Match}} \gets \emptyset\), \(\Delta \gets \emptyset\) \\

\ForEach{\(p \in V\)}{
  \ForEach{time point \(t\)}{
    Set \(\textit{Match} \gets \text{True}\), \(\delta \gets 0\) \\

    Let base condition code \(i_0 \gets 1\) (fixed reference condition) \\

    \ForEach{\(r \in \{1,2\}\)}{

      \For{\(i = 2 \text{ to } 7\)}{
        Let  \(f_{i} \gets \mathcal{F}(i)\) \\
        Compute fold-change: \(d_{(i, p, t)}^{(r)} \gets x_{(i,p,t)}^{(r)} / x_{(i_0, p, t)}^{(r)}\) \\
        \If{\(|d_{(i, p, t)}^{(r)} - f_{i}| > \epsilon\)}{
          \(\textit{Match} \gets \text{False}\); \textbf{break}
        }
        \(\delta \gets \delta + |d_{(i, p, t)}^{(r)} - f_{i}|\)
      }
    }

    \If{\(\text{Match}\)}{
      Add \(t\) to \(\mathcal{V}_{\mathrm{Match}}[p]\) \\
      Store deviation: \(\Delta[(p,t)] \gets \delta\)
    }
  }
}

\textbf{Step 2: Select Best Gene-Time Pair} \\
\((p^*, t^*) \gets \arg\min_{(p,t) \in \Delta} \Delta[(p,t)]\)
}
\end{algorithm}

\suppsubsection{Sub-GRNN Search Algorithm for Mathematical Calculation Tasks}\label{sec:calculation}

To search for sub-GRNNs whose output gene expression profiles match predefined mathematical calculation patterns, we use a fold-change matching algorithm (Algorithm~\ref{alg:calculation}). 

We  compute fold-changes relative to the baseline condition ($i = \text{base}$) and identify genes matching expected patterns across all input codes ($i = 1$ to $6$) in both replicates.
The algorithm checks each gene \( p \) at each time point \( t \) in the GRNN, comparing its observed fold-changes across inputs (\( i = 1,\dots,6 \)) with the expected fold-change pattern \( \mathcal{F}(i) \), evaluated independently in both replicates.  
A tolerance threshold \(\epsilon\) is applied to accommodate measurement variability. Only the output-gene and the time point that match the expected fold-change pattern across all input codes and replicates are retained as matching output-genes \((p,t) \in \mathcal{V}_{\mathrm{Match}}\).
From \(\mathcal{V}_{\mathrm{Match}}\) the best matching output gene and time point \((p^*, t^*)\) is the pair with the lowest fold-change deviation from the expected fold change.
This output gene is then used to trace upstream regulatory pathways and extract the sub-GRNN.


\suppsubsection{Sub-GRNN Search Algorithm for Classfication Tasks}\label{sec:classification}

For classification problems, we identify genes whose expression matches a predefined binary pattern  across all input codes ($i=1,2,..., 7$) in both replicates using a two-stage search algorithm (Algorithm ~\ref{alg:classification}, ~\ref{alg:classification_step2}). 


In the first stage (Algorithm~\ref{alg:classification}), each gene \( p \in V \) is evaluated at each time point to test whether its expression can separate the input conditions into two classes using the decision threshold. In this step, we search for genes whose expression pattern changes consistently with the input codes. When a chemical compound corresponding to a given input code is added to the cells, the correct classification requires the gene to be expressed above the decision threshold, while for non-matching input codes its expression should remain below the threshold. 
Only the output genes and the corresponding time that maintain this class separation consistently across replicates and input conditions are retained.


In the second stage (Algorithm~\ref{alg:classification_step2}), we rank the output genes selected in the first stage by how well they separate the two classes. For each gene, we measure the distance between the expression values just above the threshold and those just below it. The gene and time point with the largest separation distance are chosen, since a wider margin gives more reliable classification.  
 Finally, the best-matching output-gene and the time point pair \((p^\star, t^\star)\) is selected along with decision thresholds \(\hat{\theta}_{p^\star}^{(1)}\)  and \(\hat{\theta}_{p^\star}^{(2)}\) for both replicates.

\suppsubsection{Sub-GRNN Search for the Problem What is the Collatz Step Count for $i$ ?}
\label{sec:collatz-method}

To identify gene expression patterns corresponding to Collatz step counts, we implement a binary vector matching algorithm (Algorithm~\ref{alg:collatz-step-selection}). 

For each gene \( p \in V \) at each time point \( t \), the algorithm calculates a decision threshold by finding the largest gap between two neighboring expression levels. This threshold splits the gene’s expression profile into two states. The values above the threshold are assigned 1, and values below it are assigned 0. The resulting binary pattern is then compared to the expected Collatz step-count binary pattern. The output-genes and the corresponding time point is selected as the best-matching output-genes if its binary pattern matches the Collatz sequence in both replicates. Among all best-matching output-genes, the final output gene and time point \((p^*, t^*)\) are chosen as the one with the largest distance between the expression values just above and just below the threshold, ensuring reliable binary encoding.

We search for genes whose expression patterns follow a predefined binary pattern across input codes ($i = 1 \dots 7$) in both replicates. At each time point, a threshold is set as the midpoint of the largest adjacent expression gap, and the best-matching gene is chosen by maximizing the separation between expression values above and below this threshold. Using this approach, the algorithm identifies five output genes (\textit{b1093}, \textit{b1779}, \textit{b3165}, \textit{b3123}, and \textit{b0759}) whose thresholded expression patterns at \(t = 6\) hours precisely reconstruct the Collatz step count using binary weights \(2^0\) to \(2^4\).  
For example, input code \(i = 2\) yields an output of 1, and only \textit{b1093} (i.e., \(2^0\)) is expressed above threshold, with all others remaining below. At \(i = 3\), the output is 7, and genes \textit{b1093}, \textit{b1779}, and \textit{b3165} are all strongly activated, matching the binary pattern \(111\).

\begin{figure}[t!] 
\centering
\includegraphics[width=0.7\linewidth]{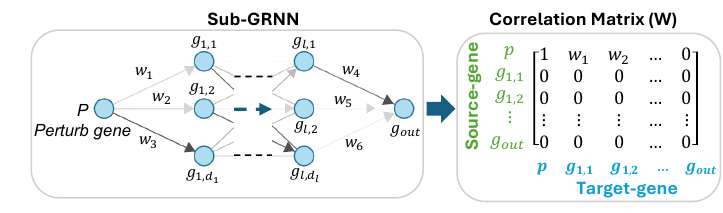} 
\vspace{-1.5em}
\caption{\textbf{The correlation matrix of the sub-GRNN.} The perturbation is applied at gene \(p\) and propagates layer by layer through intermediate nodes \(g_{1,1}, g_{1,2}, \dots, g_{l,d_l}\) to the output gene \(g_{\text{out}}\). Directed edges are weighted by \(w_i\), and the network is represented in the correlation matrix \(\mathbf{W}\), where rows correspond to source genes and columns to target genes. Each entry \(W_{ij}\) denotes the regulatory influence of source gene \(i\) on target gene \(j\), with \(W_{pp} = 1\) marking the perturbed gene \(p\).   
}
\vspace{-1.5em}
\label{fig:correlation_matrix}
\end{figure}

\suppsection{Perturbation Analysis}
This section outlines the algorithms used in perturbation analysis, including gene-wise perturbation, collective perturbation, and Lyapunov-based stability methods. 

\suppsubsection{Gene-wise Perturbation Analysis}
\label{sec:perturbation-analysis}

After identifying sub-GRNNs whose expression dynamics match the task-specific target patterns, we evaluate their computing stability and reliability using a gene-wise perturbation analysis (Algorithm~\ref{alg:nodewise_perturbation}). 
This analysis tests how perturbing a single hidden-layer gene \(p\) affects the the output gene expression. The key idea is to use a correlation-weighted adjacency matrix \(W\), which describes how strongly each gene influences its neighbors.
Each row of \(\mathbf{W}\) indexes a source gene and each column a target gene, with \(W_{ij}\) quantifying the influence from source \(i\) to target \(j\).  Under a perturbation of gene \(p\), we set \(W_{pp}=1\) and \(W_{jj}=0\) for all \(j\neq p\), so that \(\mathbf{W}\) encodes the directed propagation of the perturbation from \(p\) through hidden-layer genes to the output gene \(g_{\text{out}}\) (Figure~\ref{fig:correlation_matrix}).
First, a small random perturbation \(\mathbf{u}_t\) is generated for gene \(p\). This perturbation is sampled from a Gaussian distribution with zero mean, then scaled by a fold-change factor \(\alpha\) and a noise variance \(\sigma^2\) to control its strength. The perturbation is then propagated downstream through the network according to the correlation weights in \(W[p,:]\). The strongly correlated edges propagate more of the perturbation, than weakly correlated edges.  
At each time point, the propagated perturbations are added to the original unperturbed expression levels, producing perturbed expression levels \(\tilde{\mathcal{X}}^{(\alpha)}_{c,r}\) for each condition and replicate. 


For calculation tasks, we compute fold-change deviation after perturbation and calculate the coefficient of determination ($R^2$) using Equation 5 in the main text. 
We define gene's Criticality 
as the ratio of its outward degree centrality to  mean \(R^2\) across perturbation levels (Equation 7 in the main text). The critical genes 
are those whose perturbation strongly affects output expression, with higher criticality indicating greater disruption to the output-gene expression.
For classification tasks, we compute the Hamming distance between the perturbed and unperturbed binary outputs across input codes (Equation 8 in the main text). Classification-based criticality is the product of centrality and total Hamming distance across perturbation levels (Equation 9 in the main text).

\begin{algorithm}[H]
\caption{Sub-GRNN Search Algorithm for Classification Problems: Identify Genes that Matches Given Classification Pattern}
\label{alg:classification}
\KwIn{
\begin{itemize}
  \item Gene expression levels \(x^{(r)}_{(i, p, t)}\) of $p^{th}$ gene corresponding to  $i^{th} \in \mathcal{I}$ input code  at time $t=6, 30, 70$ hours,  for replicate \(r = 1,2\).
  \item  Gene set \(V\) in the GRN.
  \item Input codes  \(\mathcal{I} = \{1, 2, \ldots, 7\}\).
  \item Target classification pattern \(\mathcal{F} \subseteq \mathbb{Z}^+\).
\end{itemize}
}
\KwOut{Mathing output-gene  \(p^\star\),  with time point \(t^\star\) and midpoint thresholds \(\hat{\theta}_{p^\star}^{(r)}\)}

\Begin{
\textbf{Step 1: Identify Output-genes that Match the  Pattern \(\mathcal{P}\)}\\
Initialize \(\mathcal{V}_{\mathrm{Match}} \gets \emptyset\), \(\mathcal{V}_{\mathrm{used}} \gets \emptyset\)\\

\ForEach{\(p \in V\)}{
  \If{\(p \in \mathcal{V}_{\mathrm{used}}\) or  \(x^{(r)}_{(i,p,t)} \neq \emptyset\)}{ continue }

  Compute gene-specific threshold \(\theta_p \gets \frac{1}{|\mathcal{I}|} \sum x^{(r)}_{(i, p, t)}\), the mean expression level of \(p\) across all input code.\\
  Set \(T_p \gets \emptyset\) \tcp*{Time points satisfying classification pattern}

  \ForEach{time point \(t\)}{
    Set \(\mathcal{V}_{\mathrm{Match}} \gets \text{True}\)\\

    \ForEach{replicate $r$ and expected label $\mathcal{F}$}{
      \If{\(p \notin x^{(r)}_{(i, p, t)}\)}{ \(\mathcal{V}_{\mathrm{Match}} \gets \text{False}\); break }


      \eIf{\(I_i \in \mathcal{F}\)}{
        \If{\(x^{(r)}_{(i, p, t)} \leq \theta_p\)}{ \(\mathcal{V}_{\mathrm{Match}} \gets \text{False}\); break }
      }{
        \If{\(x^{(r)}_{(i, p, t)} > \theta_p\)}{ \(\mathcal{V}_{\mathrm{Match}} \gets \text{False}\); break }
      }
    }

    \If{\(\mathcal{V}_{\mathrm{Match}} \neq \emptyset\)}{ \(T_p \gets T_p \cup \{t\}\) }
  }

  \If{\(T_p \neq \emptyset\)}{
    \(\mathcal{V}_{\mathrm{Match}}[p] \gets T_p\); \(\mathcal{V}_{\mathrm{used}} \gets \mathcal{V}_{\mathrm{used}} \cup \{p\}\)
  }
}
}
\end{algorithm}

\begin{algorithm}[H]
\caption{Sub-GRNN Search Algorithm for Classification Problems: Determine the Best Gene for the given Classification problem}
\label{alg:classification_step2}

\textbf{Step 2: Evaluate Expression Separation and Rank Genes}\\
Initialize: Best score \(s^\star \gets \infty\), best output gene \(p^\star \gets \text{None}\)

\ForEach{\(p \in \mathcal{V}_{\mathrm{Match}}\)}{
  Set \(T_p \gets \mathcal{V}_{\mathrm{Match}}[p]\) and
  initialize thresholds \(\theta_p^{(r)}\) as mean expression of \(p\) in \(x^{(r)}_{(i, p, t)}\) for replicates $r=1,2$.

  \ForEach{time point \(t \in T_p\)}{
    Set: \(\delta^{(r)}_{\text{above}}, \delta^{(r)}_{\text{below}} \gets \infty\)\\

    \ForEach{$i\in \mathcal{I}$ , replicate $r$}{
      \If{\(x^{(r)}_{(i, p, t)} \geq \theta_p^{(r)}\)}{
        \(\delta \gets x^{(r)}_{(i, p, t)} - \theta_p^{(r)}\)\\
        \If{\(\delta < \delta^{(r)}_{\text{above}}\)}{
          \(\delta^{(r)}_{\text{above}} \gets \delta\), store \(x^{(r)}_{(i, p, t)}\) as closest above
        }
      }\Else{
        \(\delta \gets \theta_p^{(r)} - x^{(r)}_{(i, p, t)}\)\\
        \If{\(\delta < \delta^{(r)}_{\text{below}}\)}{
          \(\delta^{(r)}_{\text{below}} \gets \delta\), store \(x^{(r)}_{(i, p, t)}\) as closest below
        }
      }
    }

      }
    }

  Compute midpoint thresholds using, 
  \(\hat{\theta}_p^{(r)} \gets \frac{1}{2}(\text{closest above} + \text{closest below})\)

  Total score: \(s_p \gets \sum_r  \delta^{(r)}_{\text{above}} + \delta^{(r)}_{\text{below}}\)\\
  \If{\(s_p < s^\star\)}{
     \(s^\star \gets s_p\), \(p^\star \gets p\), \(t^\star \gets t\)
    \(\hat{\theta}_{p^\star}^{(r)} \gets \hat{\theta}_p^{(r)}\)
  }
\end{algorithm}

\begin{algorithm}[H]
\caption{Sub-GRNN Search Algorithm for Collatz Step Count}
\label{alg:collatz-step-selection}

\KwIn{
\begin{itemize}
  \item Gene expression levels \(x^{(r)}_{(i, p, t)}\) of $p^{th}$ gene corresponding to  $i^{th} \in \mathcal{I}$ input code  at time $t=6,30,70$ hours  for replicate \(r = 1,2\).
  \item Gene set \(V\) in the GRN.
  \item Expected binary pattern: \(\mathbf{b} \in \{0,1\}^I\)
\end{itemize}
}

\KwOut{
The best-matching genes \((p^*)\) with matched binary patterns at time point \((t^*)\), and threshold midpoint \((\mu^*)\) maximizing expression separation.}

\Begin{
Initialize: \(\mathcal{V}_{\mathrm{Match}} \gets \emptyset\), \(\Delta \gets \emptyset\), \(p^* \gets \emptyset\), \(t^* \gets \emptyset\), \(\mu^* \gets \emptyset\), \(n_{\max} \gets 0\) \\

\ForEach{gene \(p \in V\)}{
  \ForEach{time point \(t \)}{
    \ForEach{replicate \(r\)}{
      Compute threshold as \(\theta^{(r)} \gets \arg\max_{\text{adjacent } i} |x^{(r)}_{(i+1, p, t)} - x^{(r)}_{(i, p, t)}|\) \\(midpoint of max adjacent difference) \\
      Generate binary vector \(\hat{\mathbf{b}}^{(r)} \in \{0,1\}\): \(\hat{b}^{(r)}_i = 1\) if \(x^{(r)}_{(i, p, t)} > \theta^{(r)}\), else 0
    }

    \If{\(\hat{\mathbf{b}}^{(1)} = \mathbf{b} \land \hat{\mathbf{b}}^{(2)} = \mathbf{b}\)}{
      Add time point \(t\) to \(\mathcal{V}_{\mathrm{Match}}[p]\) 

      Find: \(x^+ \gets \min\{x \in x^{(r)}_{(i, p, t)} : x \geq \theta^{(r)}\}, \quad x^- \gets \max\{x \in x^{(r)}_{(i, p, t)} : x < \theta^{(r)}\}\) 
      Compute separation \(n \gets x^+ - x^-\) and midpoint \(\mu = \frac{x^+ + x^-}{2}\) \\

      \If{\(n > n_{\max}\)}{
        \(p^* \gets p\), \(t^* \gets t\), \(\mu^* \gets \mu\), \(n_{\max} \gets n\), \(\Delta[p] \gets n\)
      }
    }
  }
}

Sort \(\mathcal{V}_{\mathrm{Match}}\) genes by \(\Delta[p]\) (descending) and return top genes with the time point. Also, return \((p^*, t^*, \mu^*)\)
}
\end{algorithm}

\begin{algorithm}[H]
\caption{Gene-wise Perturbation Analysis.}
\label{alg:nodewise_perturbation}

\KwIn{
\begin{itemize}
\item Gene set \(V\) in the GRN.
  \item Subnetwork edges \(\mathcal{E} \subseteq V \times V\), edge correlations \(\rho: \mathcal{E} \rightarrow [-1,1] \cup \{\emptyset\}\)
  \item Unperturbed  Gene expression levels \(x^{(r)}_{(i, p, t)}\) of $p^{th}$ gene corresponding to  $i^{th} \in \mathcal{I}$ input code  at time $t=6,30,70$ hours,  for replicate \(r = 1,2\).
  \item Perturbation gene \(p \in V\), max depth \(D_{\max}\), fold factors \(\boldsymbol{\alpha} = \{\alpha_1, \dots, \alpha_k\}\), noise variance \(\sigma^2\)
  \item \(\Pi_{(i,r)}(j)\) denotes the cumulative product of correlation coefficients along any path in the GRN that reaches gene \(j\) under input code \(i\) and replicate \(r\).
\end{itemize}
}
\KwOut{
Perturbed gene expression matrices \(\tilde{x}^{(r)}_{(i, p, t)}\)
}

\Begin{

\textbf{Step 1: Assign Missing Correlations} \\
\ForEach{\((i,j) \in \mathcal{E}\)}{
  \If{\(\rho(i,j) = \emptyset\)}{ \(\rho(i,j) \sim \mathcal{U}(-1,1)\) }
}

\textbf{Step 2: Compute Gene-wise Correlation Matrix \(W\)} \\
\ForEach{Input code  \(i\), replicate \(r\)}{
  Initialize \(\Pi_{(i,r)}(j) \gets \emptyset\) for all \(j \in V\) \\
  Initialize queue \(Q \gets [(p,1,0)]\)

  \While{\(Q \neq \emptyset\)}{
    Dequeue \((v, \pi, D)\); \If{\(D \geq D_{\max}\)}{continue}
    \ForEach{\((v,v') \in \mathcal{E}\)}{
      \(\pi' \gets \pi \cdot \rho(v,v')\); append \(\pi'\) to \(\Pi_{(i,r)}(v')\); enqueue \((v', \pi', d+1)\)
    }
  }

  Construct \(W_{(i,r)}[p,:]\) by: \\
  \(W_{(i,r)}[p,v] \gets \text{normalize}(\sum \Pi_{(i,r)}(v))\), with \(W_{(i,r)}[p,p] \gets 1\)
}

\textbf{Step 3: Add Propogated Perturbation to Unperturbed Gene Expression Levels.} \\
For each \(i\), compute: \\
\(x_{\min}(p) \gets \min_t x^{(r)}_{(i, p, t)},\quad x_{\max}(p) \gets \max_t x^{(r)}_{(i, p, t)}
\)

\ForEach{\(\alpha \in \boldsymbol{\alpha}\)}{
  \ForEach{\((i,r)\)}{
    Let \(W = W_{(i,r)}\)

    \ForEach{time \(t\)}{
      Sample perturbation input: 
      \(\mathbf{u}_t \sim \mathcal{N}(0,1) \cdot \sigma^2 \cdot \mathcal{U}(x_{\min}(p)(1-\alpha), x_{\max}(p)(1+\alpha))\) 

      Propagate: \\
      \(\tilde{x}^{(r)}_{i, p, t} \gets x^{(r)}_{(i, p, t)} + W[p,:] \cdot \mathbf{u}_t\)
    }

    Store \(\tilde{x}^{(r)}_{(i, p, t)}\)
  }
}
}
\end{algorithm}

\begin{algorithm}[H]
\caption{Collective Perturbation Analysis}
\label{alg:collective_perturbation}

\KwIn{
\begin{itemize}
  \item Subnetwork edges \(\mathcal{E} \subseteq V \times V\), where edge correlations \(\rho: \mathcal{E} \rightarrow [-1,1] \cup \{\emptyset\}\) and genes \(V\) in the GRN.
  \item Unperturbed gene expression levels \(x^{(r)}_{(i, p, t)}\) of $p^{th}$ gene corresponding to  $i^{th} \in \mathcal{I}$ input code  at time $t=6,30,70$ hours,  for replicate \(r = 1,2\).
  \item Perturbation genes \(\mathcal{P} \subseteq V\), max depth \(D_{\max}\), fold factors \(\boldsymbol{\alpha} = \{\alpha_1, \dots, \alpha_k\}\), noise variance \(\sigma^2\).
  \item \(\Pi_{(i,r)}(j)\) denotes the cumulative product of correlation coefficients along any path in the GRN that reaches gene \(j\) under input code \(i\) and replicate \(r\).
\end{itemize}
}
\KwOut{
Perturbed gene expression matrices \(\tilde{x}^{(r)}_{(i, p, t)}\)
}

\Begin{

\textbf{Step 1: Assign Missing Correlations} \\
\ForEach{\((i,j) \in \mathcal{E}\)}{
  \If{\(\rho(i,j) = \emptyset\)}{ \(\rho(i,j) \sim \mathcal{U}(-1,1)\) }
}

\textbf{Step 2: Compute Collective Correlation Matrix \(W\)} \\
\ForEach{Input code \(i\), replicate \(r\)}{
 Initialize \(\Pi_{(i,r)}(j) \gets \emptyset\) for all \(j \in V\)  \\
  Initialize queue \(Q \gets \{(p,1,0)\ |\ p \in \mathcal{P}\}\)

  \While{\(Q \neq \emptyset\)}{
    Dequeue \((v, \pi, D)\); \If{\(D \geq D_{\max}\)}{continue}
    \ForEach{\((v,v') \in \mathcal{E}\)}{
      \(\pi' \gets \pi \cdot \rho(v,v')\); append \(\pi'\) to \(\Pi_{(i,r)}(v')\); enqueue \((v', \pi', d+1)\)
    }
  }

  Construct \(W_{(i,r)}[\mathcal{P},:]\) by: \\
  \(W_{(i,r)}[p,v] \gets \text{normalize}(\sum \Pi_{(i,r)}(v))\) for all \(p \in \mathcal{P}\), with \(W_{(i,r)}[p,p] \gets 1\)
}

\textbf{Step 3: Apply Collective Perturbation} \\
For each \(i\), compute per-gene expression level bounds: \\
\(\forall p \in \mathcal{P},\quad x_{\min}(p) \gets \min_t x^{(r)}_{i, p, t},\quad x_{\max}(p) \gets \max_t x^{(r)}_{i, p, t}\)

\ForEach{\(\alpha \in \boldsymbol{\alpha}\)}{
  \ForEach{\((i,r)\)}{
    Let \(W = W_{(i,r)}\)

    \ForEach{time \(t\)}{
      \textbf{Sample and sum collective perturbations:} \\
      \(\mathbf{u}_t = \sum_{p \in \mathcal{P}} ( \mathcal{N}(0,1) \cdot \sigma^2 \cdot \mathcal{U}(x_{\min}(p)(1-\alpha), x_{\max}(p)(1+\alpha)) \cdot  \mathbf{e}_p) \)

      \textbf{Propagate:} \\
      \(\tilde{x}^{(r)}_{i, p, t} \gets x^{(r)}_{i, p, t} + \sum_{p \in \mathcal{P}} W[p,:] \cdot \mathbf{u}_t[p]\)
    }

    Store \(\tilde{x}^{(r)}_{i, p, t}\)
  }
}
}
\end{algorithm}

\suppsubsection{Collective Perturbation Analysis}
\label{sec:collective-perturbation}
To evaluate the impact of biologically plausible perturbations, we extend the gene-wise perturbation analysis to collective perturbations (Algorithm~\ref{alg:collective_perturbation}).

Unlike the gene-wise analysis, which perturbs one gene at a time, the collective perturbation method simulates simultaneous disruptions across a set of genes \(\mathcal{P}\). Similar to the  gene-wise perturbation analysis, the noise samples are drawn from a zero-mean Gaussian distribution scaled by the fold-change factor \(\alpha\) and noise variance \(\sigma^2\). These perturbations are then propagated through correlation-weighted matrix \(W\).  
The perturbation propagation  from all perturbed genes are summed together, creating a cumulative perturbation effect that propagated through the sub-GRNN. This cumulative perturbation effect is then added to the unperturbed expression profiles, producing perturbed output-gene expressions \(\tilde{\mathcal{X}}^{(\alpha)}_{c,r}\) under the collective perturbation. 
We evaluate sub-GRNN reliability under collective perturbation by sequentially perturbing the  \(k\)  most critical genes identified from the gene-wise analysis, starting from the most critical and adding up to \(k = 10\). The reliability is measured using \(R^2\) for calculation tasks and Hamming distance for classification tasks.  Finally, we explore how these metrics change as the most critical genes are sequentially added to the collective perturbation set.  

\suppsubsection{Assessing Perturbation Tolerance in Sub-GRNN Computing}

In this section, we describe the Lyapunov stability theorem \cite{siragy2025dynamic}, \cite{4459814} used to determine the maximum per-gene perturbation a sub-GRNN can tolerate while maintaining stable and reliable computing. 


Let \( G = (V, E) \) denote the GRN, where \( V \) is the set of \( n \) genes and \( E \) is the set of directed regulatory edges. The unperturbed gene expression levels at time \( t \) are represented by \( x^*(t) \in \mathbb{R}^n \), and the perturbed expression levels by \( x(t; \alpha, \sigma) \in \mathbb{R}^n \). Here, the parameter \( \alpha > 0 \) specifies the fold-change amplitude relative to the range between maximum and minimum unperturbed expression levels, and \( \sigma \in \mathbb{R}^n \) represents the variance of the perturbed expression.  
The correlation matrix \( \mathbf{W} \) is derived from publicly available transcriptomic datasets by computing pairwise correlations of gene expression levels, providing coefficients that capture the strength of gene–gene interactions within the GRNN.  
In the  matrix \( \mathbf{W} \in \mathbb{R}^{n \times n} \), each row \( p \) corresponds to gene \( p \) as the source-gene, each column corresponds to a target gene, and \( W_{pp} = 1 \) for all \( p \).
The perturbed gene expression levels of the sub-GRNN is given by Equation 10 in the main text.
Using Equation 10, Equation 11, and Equation 12 in the main text, we define the Lyapunov function for the mathematical calculation problems as given in Equation 13 in the main text.  
For classification problems, the Lyapunov function is given by Equation 21  in the main text.

To verify the Lyapunov conditions, we note that \( \mathcal{V}(t; \alpha, \sigma) \) is expressed as a finite sum of squared norms, which are always non-negative, and therefore \( \mathcal{V}(t; \alpha, \sigma) \geq 0 \) for all \( t \), establishing positive semi-definiteness.  Furthermore, \( \mathcal{V}(t; \alpha, \sigma) = 0 \) holds if and only if \( \delta x_p = 0 \) and \( \delta x_q = 0 \) for all \( (p,q) \in E \).  

To determine the maximum perturbation level that a sub-GRNN can tolerate while maintaining stable and reliable computing, we analyze the Lyapunov derivative in Equation 28 in the main text. When \( \tfrac{dV}{ds} < 0 \), the sub-GRNN moves toward stability, meaning it remains in a steady state with output expression close to the expected pattern. When \( \tfrac{dV}{ds} \geq 0 \), the system instead tends toward instability.  Then we solve \( \tfrac{dV}{ds} = 0 \) (Equation 28 in the main text) to identify the maximum perturbation level that the sub-GRNN can tolerate.  
It is true that $\frac{\sigma(s)}{\alpha(s)} \neq 0.$
and to solve \( \frac{d\mathcal{V}}{ds} = 0 \), we require:
\begin{equation}
    2l - \frac{k \sigma(s)}{\alpha(s)^3 \left\| \Delta_{q} \right\|} = 0.
\end{equation}

By substituting the expressions for \( \sigma(s) \) and \( \alpha(s) \) from Equation 29 ($\alpha(s)=\alpha_0 + k s$) and Equation 30 ($\sigma(s)= \sigma_0 + ls$) in the main text, we obtain:

\begin{align}
\sigma_0 + ls &= \frac{2l}{k} (\alpha_0 + ks)^3 \left\| \Delta_{q} \right\|\\
     2lk^3 s^3 + 6lk^2\alpha_0 s^2 + &(6lk\alpha_0^2 - kl) s + (2l\alpha_0^3 - k\sigma_0) =0.
\end{align}

Therefore, the perturbation level at which instability emerges due to increasing noise on a perturbed gene is determined by the roots of the following expression:
\begin{align}
s_1 = & -1.23 \left(
\sqrt{\frac{0.03}{k^4 l^2} - \frac{0.02}{k^5 l^2} + \frac{1}{k^6}}
- \frac{0.12}{k^2 l} + \frac{0.12}{k^3}
\right)^{1/3} \notag \\
& - \frac{0.1}{k}
- \frac{1.23}{k^2 \left(
\sqrt{\frac{0.03}{k^4 l^2} - \frac{0.02}{k^5 l^2} + \frac{1}{k^6}}
- \frac{0.12}{k^2 l} + \frac{0.12}{k^3}
\right)^{1/3}}.
\end{align}

We aim to establish that \( s_1 > 0 \). To simplify the expression, let us define:
\begin{equation}
A = 
\sqrt{\frac{0.03}{k^4 l^2} - \frac{0.02}{k^5 l^2} + \frac{1}{k^6}}
- \frac{0.12}{k^2 l} + \frac{0.12}{k^3}.
\end{equation}

Then the expression for \( s_1 \) can be rewritten as:
\begin{equation}
s_1 = -1.23 \left( A^{1/3} + \frac{1}{k^2  A^{1/3}} \right) - \frac{0.1}{k\left\| \Delta_{q} \right\|}.
\end{equation}

To prove that \( s_1 > 0 \), it is sufficient to demonstrate that \( A < 0 \). Since we assume that the gene expression levels are real-valued (i.e., \( s_1 \in \mathbb{R} \)), the square root must be non-negative:
\begin{equation}
\sqrt{\frac{0.03}{k^4 l^2} - \frac{0.02}{k^5 l^2} + \frac{1}{k^6}} > 0, \quad \text{for all } k, l > 0.
\end{equation}

Now observe that:
\[
\frac{0.12}{k^3} - \frac{0.12}{k^2 l} > \sqrt{\frac{0.03}{k^4 l^2} - \frac{0.02}{k^5 l^2} + \frac{1}{k^6}}.
\]
This inequality implies that:
\[
A = \sqrt{\cdots} - \frac{0.12}{k^2 l} + \frac{0.12}{k^3} < 0.
\]

Since \( A < 0 \), its cube root \( A^{1/3} \) is also negative. Therefore, each of the terms \( -1.23 A^{1/3} \) and \( -1.23/(k^2 A^{1/3}) \) becomes positive. The only negative term in \( s_1 \) is \( -0.1/k \), which is small for \( k > 0 \).
Thus, 
$s_1 > 0 \quad \text{for all } k, l > 0 \text{ satisfying } A < 0.$

The second root is found to be negative:
\[
s_2 = -\frac{0.1}{l},
\]
since \( \frac{0.1}{l} > 0 \) for all \( l > 0 \). Thus, \( s_2 < 0 \), and this root lies strictly in the left-half complex plane.

\medskip

Next, consider the remaining roots \( s_3 \) and \( s_4 \), which are complex conjugates. Since these are nonreal and the second root is negative, only the real positive root is retained under the assumption \( s > 0 \). The complex roots are expressed as:

\begin{align}
s_3 = & \, 0.0052 (1 - 1.73i) \Bigg(
\frac{
5k^2 \left( 
A^{2/3} (1 + 1.73i)^2 
\right)
- 4.76k A^{1/3} (1 + 1.73i) + 251.98
}
{
k^2 A^{1/3}
} \Bigg), \\
s_4 = & \, 0.0052 (1 + 1.73i) \Bigg(
\frac{
5k^2 \left( 
A^{2/3} (1 - 1.73i)^2 
\right)
- 4.76k A^{1/3} (1 - 1.73i) + 251.98
}
{
k^2 A^{1/3}
} \Bigg),
\end{align}
where
\[
A = \frac{\sqrt{-2000l^2 + 81(k - l)^2}}{k^3 l} - 9k + 9l.
\]

\suppsection{Additional Results} \label{sec:Additional_results}
This section provides additional results that support the analyses of mathematical calculation and classification tasks described in this study.

\suppsubsection{What is the \( i^{\text{th}} \) Fibonacci number? }
\label{sec:fibonacci_supp}

\begin{figure}[t!] 
\centering
\includegraphics[width=\linewidth]{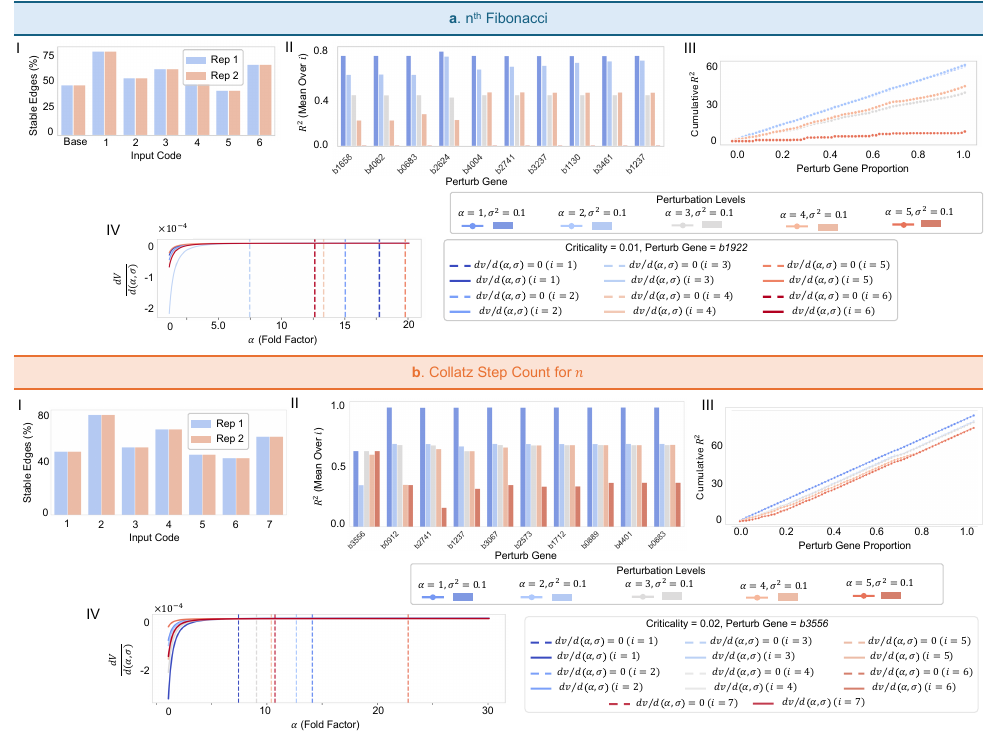} 
\vspace{-2.5em}
\caption{\textbf{Additional analyses for What is the ith Fibonacci number? (a) and Collatz step counts for $i$ (b).} 
\textbf{a,} Additional analysis of the sub-GRNN $i^{\text{th}}$ Fibonacci number output pattern. 
\textbf{I,} Proportion of stable  edges across input codes, estimated using publicly available transcriptomic datasets. 
\textbf{II,} Mean coefficient of determination $R^2$ across inputs for gene-wise perturbations. 
\textbf{III,} Cumulative $R^2$ across perturb gene proportion of gene-wise perturbation. 
\textbf{IV,} Lyapunov stability analysis for the least critical gene. 
\textbf{b,} Additional analysis of the Collatz step counts for $i$ sub-GRNN. 
\textbf{I,}  Proportion of stable  edges across input codes.
\textbf{II,}  Mean coefficient of determination $R^2$ across inputs for gene-wise perturbations.
\textbf{III,} Cumulative $R^2$ across perturb gene proportion of gene-wise perturbation.
\textbf{IV,} Lyapunov stability analysis for the least critical gene.
}

\vspace{-1em}
\label{fig:collatz_Sup}
\end{figure}

We evaluate the edge stability of the \( i^{\text{th}} \) Fibonacci number sub-GRNN  by quantifying the proportion of stable gene–gene interactions under each input condition. Using publicly available transcriptomic datasets, we found that approximately 30\% of GRN edges are consistent across diverse conditions. Figure \ref{fig:collatz_Sup}a(I) shows the fraction of these stable edges retained in the \( i^{\text{th}} \) Fibonacci number sub-GRNN. Input codes \textit{Base}, \textit{1}, and \textit{3} show the highest edge stability across replicates, suggesting that these chemical compounds activate more conserved regulatory pathways. In contrast, input codes \textit{4} and \textit{5} induce more dynamic  edges. This behavior occurs because input codes $i=$ Base, 1, and 3 corresponding to Glucose rely on stable and well-conserved metabolic pathways, while Nitrate and MnCl$_2$ input codes $i=4,5$ activate stress-related pathways.

The gene-wise perturbation (Figure~\ref{fig:collatz_Sup}a(II)) shows that most genes maintain high coefficient of determination \(R^2\) under low perturbation levels, but \(R^2\) declines more substantially as perturbation strength increases. In particular, perturbing genes such as \textit{b1658} and \textit{b4062} causes pronounced reductions in \(R^2\), indicating their high sensitivity, whereas several other genes retain relatively higher \(R^2\) values, demonstrating greater robustness to perturbations. 
To capture how influence accumulates across individual genes, we compute cumulative  coefficient of determination $R^2$ of gene-wise perturbations in decreasing order of criticality \cite{Makse2024}. Figure~\ref{fig:collatz_Sup}a(III) shows that under mild perturbation ($\alpha = 1$), coefficient of determination $R^2$ increases steadily across the ranked list. The corresponding perturbation levels for $\alpha$ are indicated in the legend below the figure. At higher perturbation levels ($\alpha = 4$ and $5$), the curve flattens early, revealing that only a small proportion of most critical genes accounts for most of the cumulative $R^2$. 
This behavior stems from the presence of a narrow set of dominant perturbed genes that influence the $i^{th}$ Fibonacci number sub-GRNN.

We apply Lyapunov-based analysis to the least critical gene identified from gene-wise perturbation to determine input-specific perturbation thresholds at which the Fibonacci sub-GRNN transitions from stable to unstable computing. Figure~\ref{fig:collatz_Sup}a(IV) shows the Lyapunov derivative $\frac{dV}{d(\alpha, \sigma)}$ for gene \textit{b1922} across all input codes. Although this gene exhibits low criticality, several inputs particularly input codes \textit{i = 3}, \textit{5}, and \textit{6}, reach instability at relatively low perturbation levels ($\alpha \approx 10$–12), while other input codes such as \textit{i = 1} and \textit{2} remain stable even at $\alpha > 18$. Among all input codes, \textit{i = 3} defines the most sensitive input code, setting the lower bound on tolerable perturbation. This behavior arises because the input chemical compounds corresponding to the input codes \textit{i = 3} drive stronger perturbation propagation through the sub-GRNN,  leading to pronounced changes in output-gene expression.  


\suppsubsection{What is the Collatz Step Count for $i$ ?}
The correlation analysis shows that the Collatz step count sub-GRNN contains input-specific differences in edge stability. Input codes \textit{2}, \textit{4}, and \textit{7} preserve the highest proportion of stable regulatory edges (Figure~\ref{fig:collatz_Sup}b(I)), reflecting consistent pathway activation. In contrast, input codes \textit{3} and \textit{6} exhibit reduced edge stability.  
The gene-wise perturbation analysis (Figure~\ref{fig:collatz_Sup}b(II)) reveals sharp declines in coefficient of determination \(R^2\) for perturbation-sensitive genes such as \textit{b0912} and \textit{b3556}, indicating their high sensitivity to perturbation.
The cumulative coefficient of determination $R^2$ distribution  (Figure~\ref{fig:collatz_Sup}b(III)) increase almost linearly with the proportion of perturbed genes, with steeper growth at lower perturbation levels ($\alpha = 1,2,3$) and a noticeable shift to lower slopes under stronger perturbations ($\alpha = 4,5$).  

Lyapunov stability  analysis of the least critical gene \textit{b3556} (Figure~\ref{fig:collatz_Sup}b(IV)) highlights input-specific perturbation sensitivity. Input codes \textit{i = 3}, \textit{6}, and \textit{7} reach instability at $\alpha \approx 8$–12, while \textit{i = 1} and \textit{2} remain stable beyond $\alpha = 25$. The lowest perturbation level occurs at input \textit{7}, establishing it as the perturbation threshold for the sub-GRNN that can tolerate, while maintaining stable and reliable computing.

\suppsubsection{Is \( i \) a lucky number?}
\label{sec:lucky_supp}

\begin{figure*}[t!] 
\centering
\includegraphics[width=\linewidth]{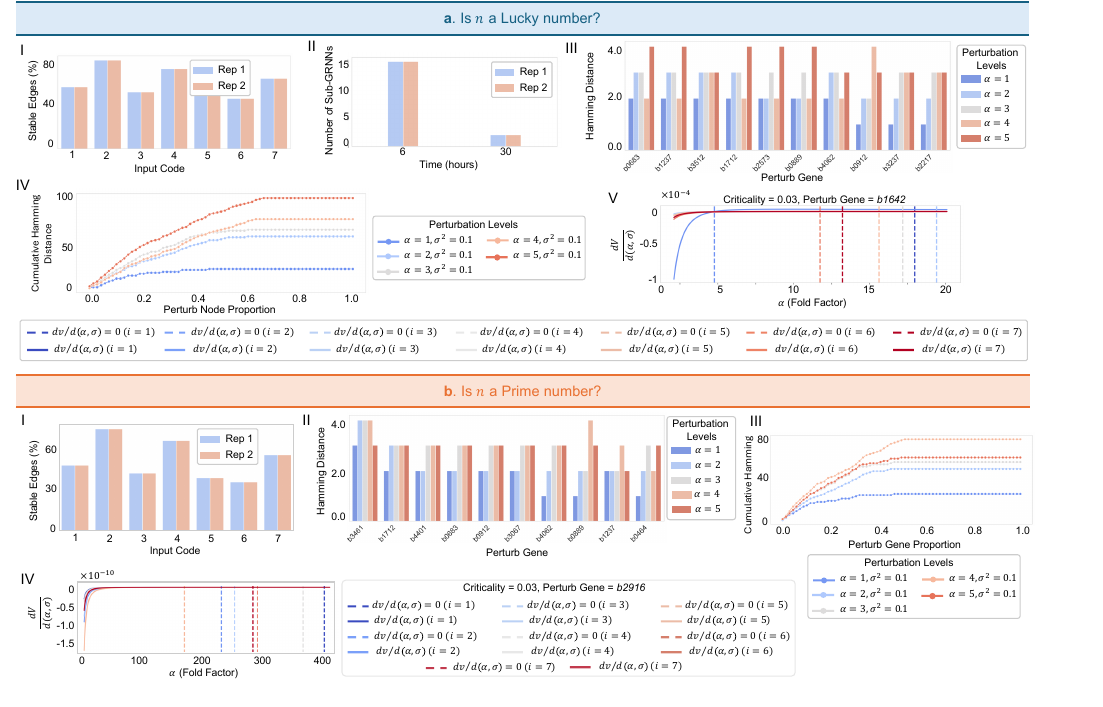} 
\vspace{-2.5em}
\caption{\textbf{GRNN-based classification of Lucky numbers (a) and Prime numbers (b).} 
\textbf{a,} Supporting analysis for the Lucky number classification. 
\textbf{I,} Proportion of stable  edges across input codes.
\textbf{II,} Number of matched sub-GRNNs over time. 
\textbf{III,} Gene-wise perturbation effects on Hamming distance for fixed noise variance \(\sigma^{2} = 0.1\) across perturbation amplitudes \(\alpha = 1, \dots, 5\).
\textbf{IV,} Cumulative Hamming distance across perturb gene proportion of gene-wise perturbation. 
\textbf{V,} Lyapunov stability analysis for the least critical gene.  
\textbf{b,} Supporting analysis for the Prime number classification. 
\textbf{I,} Input-specific edge stability in the sub-GRNN. 
\textbf{II,} Gene-wise perturbation effects on Hamming distance for fixed noise variance \(\sigma^{2} = 0.1\) across perturbation amplitudes \(\alpha = 1, \dots, 5\).
\textbf{III,} Cumulative Hamming distance across perturb gene proportion of gene-wise perturbation.
\textbf{IV,} Lyapunov stability analysis for the least critical gene \textit{b2916}.
}
\vspace{-1.3em}
\label{fig:lucky_sup}
\end{figure*}

As supporting analysis for the Lucky number classification task, we assess the stability and perturbation sensitivity of sub-GRNNs that match the expected binary expression pattern. The proportion of stable regulatory edges varies across inputs (Figure~\ref{fig:lucky_sup}a(I)), with higher proportion of stable edges obtained for input codes \textit{3}, \textit{4}, and \textit{7}, suggesting that these conditions drive more consistent regulatory behavior for this sub-GRNN. 
We then examine the number of sub-GRNNs that consist the matching classification pattern for this problem. 
The number of matched sub-GRNNs remains unchanged between 6 and 30 hours (Figure~\ref{fig:lucky_sup}a(II)).

\rev{Gene-wise perturbation results (Figure~\ref{fig:lucky_sup}a(III)) show that Hamming distance stays within a narrow range of 1–3 under mild perturbations (\(\alpha = 1, 2\)) but increases steadily with stronger perturbations, reaching the maximum at \(\alpha = 5\). This pattern highlights that perturbation of certain most critical genes disproportionately undermines classification reliability compared to others.}
Cumulative Hamming distance trends (Figure~\ref{fig:lucky_sup}a(IV)) further highlight this effect. 
Cumulative Hamming distance distribution for the gene-wise perturbation analysis (Figure~\ref{fig:lucky_sup}a(IV)) show that misclassifications rise sharply as the first 20--30\% of genes are perturbed, after which the curves gradually plateau. This indicates that a small set of core critical genes drives most of the output expression disruption, while perturbing additional genes beyond this set contributes little further change. At higher perturbation levels, the overall Hamming distance is larger, reflecting stronger error propagation and greater sensitivity of the sub-GRNN.

Lyapunov analysis of the least critical gene \textit{b1642} (Figure~\ref{fig:lucky_sup}a(V)) reveals input-dependent perturbation thresholds. Inputs codes \textit{i = 1} and \textit{7} reach instability earliest, indicating they operate closer to decision boundaries, whereas input codes such as \textit{2} and \textit{5} remain stable across a broader perturbation range. 

\suppsubsection{Is \( i \) a prime number?}
\label{sec:prime_supp}

This section presents supportive analyses of the prime number classification sub-GRNN. The correlation analysis of stable edges (Figure~\ref{fig:lucky_sup}b(I)) shows that stability varies across input codes, with \textit{2}, \textit{4}, and \textit{7} retaining a higher proportion of stable  edges. In contrast, inputs \textit{3} and \textit{6} exhibit reduced stability, suggesting that their chemical inputs modulate the sub-GRNN more strongly and introduce greater variability.

\rev{The gene-wise perturbation analysis (Figure~\ref{fig:lucky_sup}b(II)) shows that Hamming distance remains relatively stable across most genes under mild perturbations (\(\alpha = 1, 2\)) but increases sharply at stronger perturbation levels, particularly for high-critical genes such as \textit{b1712} and \textit{b3461}. This pattern underscores the selective sensitivity of the sub-GRNN, where a small subset of critical genes disproportionately influences the reliability of the output expression.}
The cumulative Hamming distance curves (Figure~\ref{fig:lucky_sup}b(III)) further confirm this nonuniform distribution of misclassifications. A small fraction of most critical genes contributes the majority of misclassification under increasing perturbation, while the remaining genes have minimal impact. C

The Lyapunov stability analysis of the least critical gene \textit{b2916} (Figure~\ref{fig:lucky_sup}b(IV)) identifies input-specific perturbation thresholds. Input codes \textit{i = 1}, \textit{2}, and \textit{4} withstand large perturbations ($\alpha > 300$), whereas input codes \textit{3}, \textit{5}, and \textit{7} become unstable at much lower levels. Among these, input code \textit{7} sets the perturbation threshold for maintaining stable and reliable computing.

\suppsubsection{Is \( i \) a Fibonacci number?}
\label{sec:fibonacci_class_supp}

\begin{figure*}[t!] 
\centering
\includegraphics[width=\linewidth]{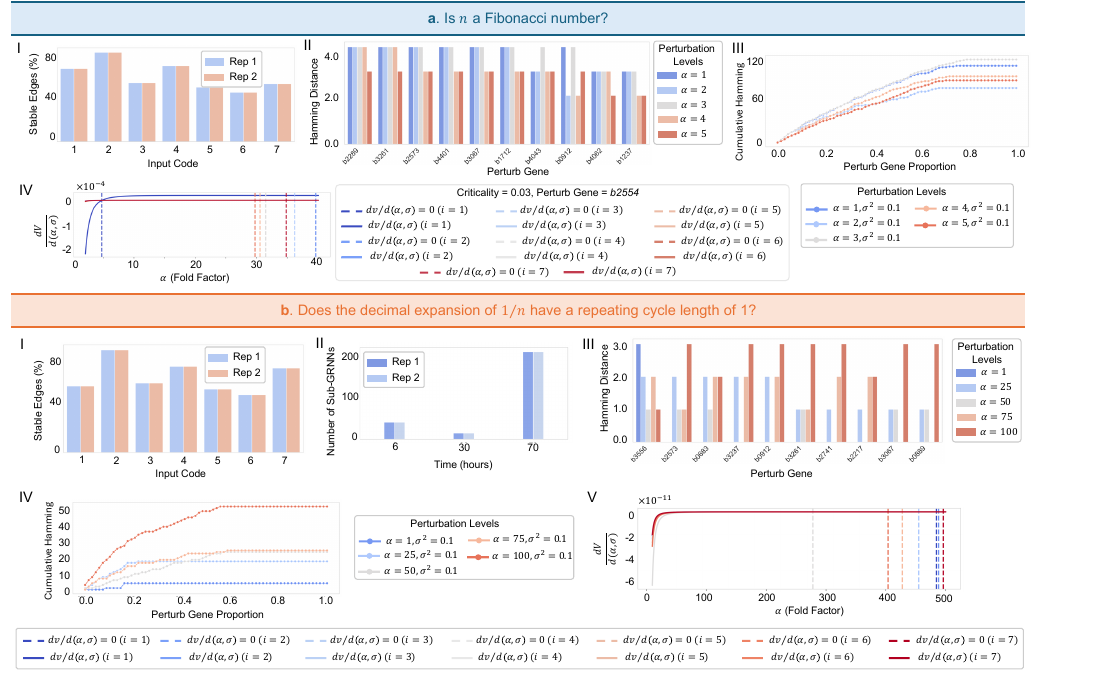} 
\vspace{-2.5em}
\caption{\textbf{GRNN-based classification of Fibonacci numbers (a) and  Does the Decimal Expansion of 1/i have a Repeating Cycle Length of 1?  (b).} 
\textbf{a,} Additional analysis of the Fibonacci number classification sub-GRNN. 
\textbf{I,} Proportion of stable gene–gene edges across input codes. 
\textbf{II,} Gene-wise perturbation effects on Hamming distance for fixed noise variance \(\sigma^{2} = 0.1\) across perturbation levels \(\alpha = 1, \dots, 5\). 
\textbf{III,} Cumulative Hamming distance across genes ranked by criticality. 
\textbf{IV,} Lyapunov stability analysis for the least critical gene.
\textbf{b,} Additional analysis of the  decimal cycle length of 1 sub-GRNN. 
\textbf{I,} Proportion of stable gene–gene edges across input codes. 
\textbf{II,} Number of matched sub-GRNNs over time. 
\textbf{III,} Gene-wise perturbation effects on Hamming distance for fixed noise variance \(\sigma^{2} = 0.1\).
\textbf{IV,} Cumulative Hamming distance across genes ranked by criticality. 
\textbf{V,} Lyapunov stability analysis across input codes for the least critical gene. 
}
\vspace{-1em}
\label{fig:decimal_sup}
\end{figure*}

In this section, we provide additional analysis of the sub-GRNN identified for  Fibonacci number classification. 
The proportion of stable  edges (Figure~\ref{fig:decimal_sup}a(I)) varies across input codes, with input codes $i=1,3,5$ exhibiting higher stability. Input codes $i=2,6$ show lower edge stability, suggesting condition-specific modulation in gene expression profiles.

\rev{The gene-wise perturbation analysis (Figure~\ref{fig:decimal_sup}a(II)) shows that Hamming distance is relatively high for most genes under mild perturbations, \(\alpha = 1, 2\) but gradually declines as perturbation levels increase (\(\alpha = 4, 5\)).}
The cumulative Hamming distance distribution (Figure~\ref{fig:decimal_sup}a(III)) shows a gradual rise followed by saturation, indicating that classification reliability is heavily dependent on a subset of hidden-layer genes. 
The Lyapunov-based stability analysis of the least critical gene \textit{b2554} (Figure~\ref{fig:decimal_sup}a(IV)) reveals that input codes \textit{i = 4}, \textit{5}, and \textit{7} reach instability earlier than the rest of input codes, with perturbation thresholds between $\alpha \approx 28$ and $34$, while input codes \textit{1} and \textit{2} remain stable throughout. 
\rev{The Lyapunov-based stability analysis of the least critical gene \textit{b2554} (Figure~\ref{fig:decimal_sup}a(IV)) shows an early instability for input \textit{i = 1} at a low perturbation threshold (\(\alpha \approx 5\)), whereas most other inputs remain stable until much larger perturbations. This pattern suggests that the \textit{i = 1} chemical condition propagates perturbations more strongly through the sub-GRNN.}


\suppsubsection{Is the repeating cycle length of \(1/i\) equal to 1?}
\label{sec:decimal_cycle_supp}

In this section, we present additional analyses of the sub-GRNNs identified for decimal cycle length classification.  
\rev{The proportion of stable edges across input codes (Figure~\ref{fig:decimal_sup}b(I)) is relatively uniform, with most input conditions maintaining between 60--80\% stable edges across replicates.  Input condition \textit{2} results in the strongest signal propagation along stable edges, while conditions \textit{5} and \textit{6} show slightly weaker propagation. The chemical inputs corresponding to input codes, such as medium glucose (\textit{3}) and low tryptophan with high lactose (\textit{6}), drive more variable regulatory activity, resulting in fewer stable edges across replicates.}  
The number of matched sub-GRNNs for this task increases significantly at 70 hours (Figure~\ref{fig:decimal_sup}b(II)), where 
it exhibits expression patterns that align with the decimal cycle length classification.  

\rev{The gene-wise perturbation analysis (Figure~\ref{fig:decimal_sup}b(III)) shows that Hamming distance remains between 0 and 2 for most genes under perturbations (\(\alpha = 1, 25, 50, 75\)), while at the highest perturbation level (\(\alpha = 100\))  it reaches the maximum value of 3. This pattern underscores the robustness of the decimal cycle length sub-GRNN, establishing it as the most perturbation-robust among the problem-specific sub-GRNNs examined in this study.}
Cumulative Hamming distance curves (Figure~\ref{fig:decimal_sup}b(IV)) show rapid early growth under higher perturbation levels, with saturation occurring after perturbing a small proportion of critical genes. 
\rev{When the least critical gene is perturbed, the Lyapunov-based stability analysis (Figure~\ref{fig:decimal_sup}b(V)) shows that all input codes exhibit large perturbation tolerance, with instability occurring only at very high perturbation levels \(\alpha> 250\) values. This demonstrates strong robustness to perturbations, although input codes \(i=3,6\) still approach instability earlier than the others.}  
 This confirms that the least critical gene has minimal impact on classification accuracy, supporting our method for identifying critical genes and their impact to to the output expression.  

\suppsubsection{Sub-GRNN Analysis for Multiplication at t = 30 Hours}
\label{sec:multiplication_supp}
In this section, we present supporting analyses of the multiplication sub-GRNN, which showed the best-matching output gene expression pattern for the multiplication factors $2,3,4$ and $5$ at \(t = 30\) hours. \rev{The perturbation results are presented according to the criticality ranking of the genes, as determined from gene-wise perturbation analysis.}

\begin{figure*}[t!] 
\centering
\includegraphics[width=\linewidth]{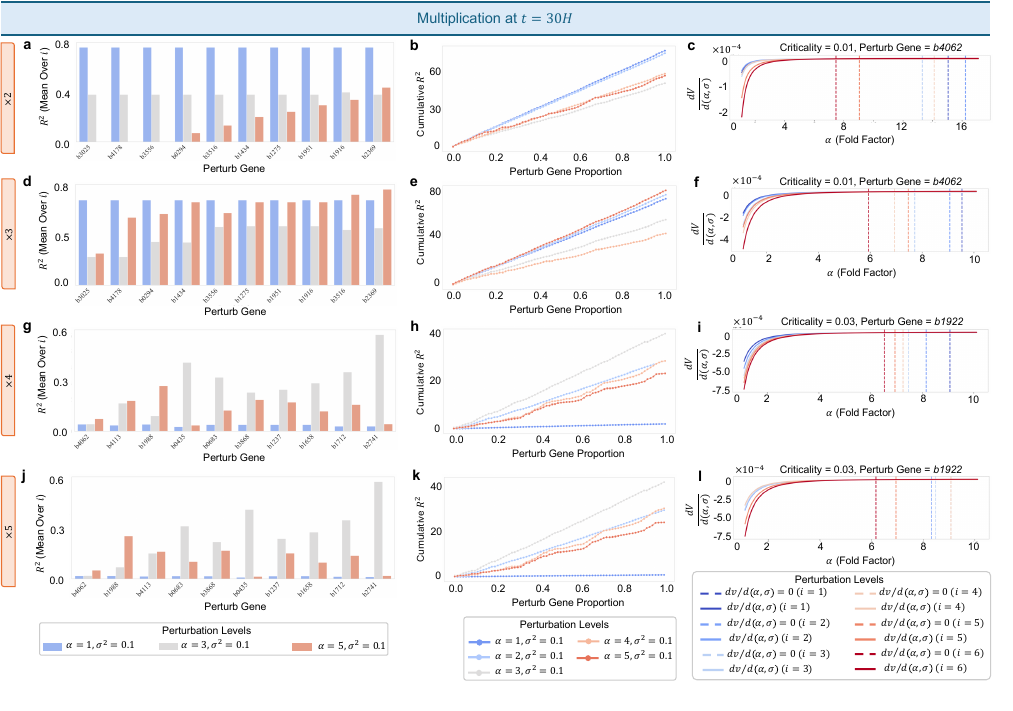} 
\vspace{-2.5em}
\caption{\textbf{GRNN-based  multiplication at $t = 30H$.} 
\textbf{a–c,} Additional analysis of the multiplication sub-GRNN for multiplication factor 2, 
\textbf{d–f,} Multiplication by 3, 
\textbf{g–i,} Multiplication by 4 and 
\textbf{j–l,} Multiplication by 5. 
Each row shows: 
\textbf{(left)} gene-wise perturbation impact on coefficient of determination $R^2$, 
\textbf{(middle)} Cumulative $R^2$ across perturb gene proportion of gene-wise perturbation, and 
\textbf{(right)}  Lyapunov stability analysis for the least critical gene.
}
\vspace{-0.2em}
\label{fig:30H_multi_sup}
\end{figure*}

\rev{For the multiplication factor 2 (Figure~\ref{fig:30H_multi_sup}a), coeffcient of determination \(R^2\) remains high across all genes at perturbation level \(\alpha = 1\), but decreases progressively under stronger perturbations, with some genes such as \textit{b3025, b4178, b3556} showing sharper drops than others.}
In contrast, genes such as \textit{b3025} and \textit{b3556} display sharp $R^2$ degradation at higher perturbation levels, reflecting their higher perturbation sensitivity and strong influence on regulatory pathways. 
The cumulative coefficient of determination $R^2$ distribution (Figure~\ref{fig:30H_multi_sup}b) increases steadily with the proportion of perturbed genes. This behavior suggests that the multiplication sub-GRNN is governed by a small set of critical genes that dominate output expression levels under perturbation.  
The Lyapunov-based stability analysis (Figure~\ref{fig:30H_multi_sup}c) confirms that the least critical gene \textit{b4062} remains stable up to perturbation levels of $\alpha > 10$ across input codes.

\rev{For the multiplication factor 3 (Figure~\ref{fig:30H_multi_sup}d), coefficient of determination \(R^2\) values are consistently high at perturbation level \(\alpha = 1\), and although they decrease at stronger  (\(\alpha = 3, 5\)), the overall decline is more gradual compared with multiplication factor 2.}  
For the multiplication factor 3 (Figure~\ref{fig:30H_multi_sup}e), the cumulative coefficient of determination $R^2$ grows steadily with the proportion of perturbed genes, but the overall curves shift downward  compared with factor 2, indicating that reliability is reduced.  The Lyapunov-based stability analysis of the least critical gene \textit{b4062} shows lower perturbation threshold near $\alpha = 6–8$, in contrast to $\alpha > 10$ for multiplication by 2. 

\rev{For the multiplication factor 4 (Figure~\ref{fig:30H_multi_sup}g), coefficient of determination \(R^2\) remains very low at perturbation level \(\alpha = 1\) for all genes. Under higher (\(\alpha = 3, 5\)), most genes show gradual increases, though the three most critical genes (\textit{b4062}, \textit{b4113}, and \textit{b1988}) stay below 0.3 across all perturbation levels.} \rev{The cumulative $R^2$ curve (Figure~\ref{fig:30H_multi_sup}h) rises gradually under higher perturbation levels but continues to increase with varying slopes.}
The Lyapunov analysis (Figure~\ref{fig:30H_multi_sup}i) of the least gene \textit{b1922} confirms this trend, showing instability emerging at $\alpha = 4$–7, much earlier than in lower multiplication factors. 

\rev{The gene-wise perturbation analysis for multiplication factor 5 (Figure~\ref{fig:30H_multi_sup}j) shows a coefficient of determination $R^2$ distribution similar to multiplication factor 4.}
\rev{The cumulative \(R^2\) curve (Figure~\ref{fig:30H_multi_sup}k) increases steadily across perturbation levels, with differences in slope reflecting the varying impact of perturbation strength.}
\rev{The Lyapunov-based stability analysis (Figure~\ref{fig:30H_multi_sup}l) of the least critical gene \textit{b1922} shows that it remains stable over a wide range of fold factors, with instability occurring only at relatively high perturbation thresholds (\(\alpha \approx 6\)–8). This confirms that least critical genes can tolerate large perturbations without strongly affecting output-gene expression.}

\begin{figure*}[t!] 
\centering
\includegraphics[width=\linewidth]{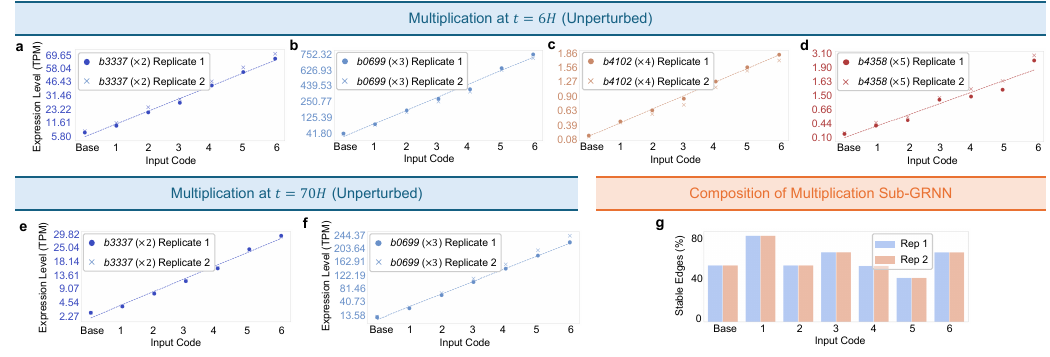} 
\vspace{-2.5em}
\caption{\textbf{ GRNN-based  multiplication at 6H and 70H, consistent with 30H.} 
\textbf{a–d,} Expression profiles of output genes matching multiplication factor  2, 3, 4, and 5 at 6H. 
\textbf{e–f,} Expression profiles of output genes matching multiplication factor 2 and 3 at 70H. 
\textbf{g,} Proportion of stable  edges of the sub-GRNN across input codes at 6H.
}
\vspace{-1em}
\label{fig:6_70H_multi_unperturb}
\end{figure*}

\suppsubsection{Sub-GRNN Analysis for Multiplication at t = 6 and 70 Hours}
\label{sec:multiplication_unperturbed}
In this section, we present additional results for the multiplication sub-GRNN identified at $t = 6$ and $70$ hours.


Using the search algorithms introduced in this study, we identified output genes \textit{b3337}, \textit{b0699}, \textit{b4102}, and \textit{b4358} for multiplication factors 2, 3, 4, and 5 at $t = 6$ hours, respectively, which match the expected patterns of this task (Figure~\ref{fig:6_70H_multi_unperturb}a-d).  
At 70H, the same genes \textit{b3337} and \textit{b0699} retain consistent fold-change behavior (Figure~\ref{fig:6_70H_multi_unperturb}e, f) for multiplication factor  two and three.  However, no matching output genes could be identified for multiplication factors 4 and 5 at $t=70$ hours.  
\rev{The correlation analysis (Figure~\ref{fig:6_70H_multi_unperturb}g) shows that the proportion of stable edges varies across input codes, with $i=1,3,6$  retaining a higher fraction of stable regulatory edges across replicates at $t=6$ hours.}  

\begin{figure*}[t!] 
\centering
\includegraphics[width=\linewidth]{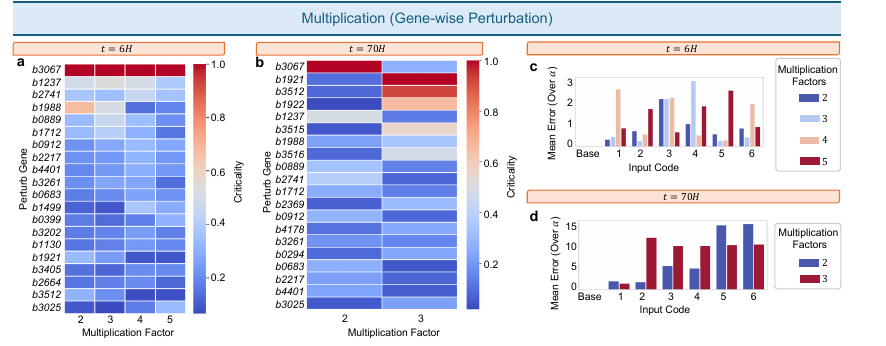} 
\vspace{-1.9em}
\caption{\textbf{The gene-wise perturbation analysis of the multiplication sub-GRNN at 6H and 70H.} 
\textbf{a–b,} The criticality distribution for top-ranked genes across multiplication factors. 
\textbf{c–d,} Mean output error across input codes under increasing perturbation levels.
}
\vspace{-1em}
\label{fig:criticality_6_70_Multi}
\end{figure*}

We further assess the computing stability and reliability of the multiplication sub-GRNN at $t=6, 70$ hours using gene-wise perturbation analysis. 
\rev{The criticality profiles (Figures~\ref{fig:criticality_6_70_Multi}a, b) reveal a consistent set of most critical genes, particularly \textit{b3067}, \textit{b1927}, and \textit{b3422}, across multiplication factors. These genes form a regulatory core whose disruption under perturbation disproportionately drives output-gene expression deviations, highlighting their key role in computing.}
Mean error over perturbation levels (Figures~\ref{fig:criticality_6_70_Multi}c, d) increases with multiplication factor, with the largest deviations observed for factors 4 and 5, especially at higher input codes.  In contrast, output-gene for the multiplication factor 2 remains the most stable, showing minimal sensitivity  perturbation at $t=70$ hours. 

\bibliography{references}